\documentclass{mn2e}
\usepackage{epsfig,amssymb,amsmath}

\newcommand{\beq}{\begin{equation}}
\newcommand{\enq}{\end{equation}}
\newcommand{\m}[1]{\boldsymbol{#1}}
\newcommand{\pa}{\partial}
\newcommand{\bra}[1]{\left#1}
\newcommand{\ket}[1]{\vphantom{\sqrt{0}}\right#1}
\newcommand{\mtext}[1]{\hspace{0.6cm}\mbox{#1}\hspace{0.6cm}}
\newcommand{\hr}{\m{\hat{r}}}
\newcommand{\hphi}{\m{\hat{\phi}}}
\newcommand{\htheta}{\m{\hat{\theta}}}
\newcommand{\hz}{\m{\hat{z}}}
\newcommand{\hvarpi}{\m{\hat{\varpi}}}

\begin{document}

\title[Stability of magnetic fields in non-barotropic stars]{Stability of magnetic fields in non-barotropic stars: an analytic treatment}
\author[T. Akg\"{u}n, A. Reisenegger, A. Mastrano and P. Marchant]
{T.~Akg\"{u}n$^{1,3}$\thanks{E-mail: akgun@astro.cornell.edu (TA); areisene@astro.puc.cl (AR)}, A.~Reisenegger$^{1}$\footnotemark[1], A.~Mastrano$^{2}$ and P.~Marchant$^{1,4}$
\\$^{1}$Departamento de Astronom\'{i}a y Astrof\'{i}sica, Facultad de F\'{i}sica, Pontificia Universidad Cat\'{o}lica de Chile,
\\ Av. Vicu\~{n}a Mackenna 4860, 782-0436 Macul, Santiago, Chile
\\$^{2}$School of Physics, University of Melbourne, Parkville, VIC 3010, Australia
\\$^{3}$Barcelona Supercomputing Center -- Centro Nacional de Supercomputaci\'{o}n, C/ Gran Capit\`{a} 2-4, Barcelona, 08034, Spain
\\$^{4}$Argelander Institut f\"{u}r Astronomie, Universit\"{a}t Bonn, Auf dem H\"{u}gel 71, D-53121, Bonn, Germany}

\onecolumn

\maketitle

\begin{abstract}
Magnetic fields in upper main-sequence stars, white dwarfs, and neutron stars are known to persist for timescales comparable to their lifetimes. From a theoretical perspective this is problematic, as it can be shown that simple magnetic field configurations are always unstable. In non-barotropic stars, stable stratification allows for a much wider range of magnetic field structures than in barotropic stars, and helps stabilize them by making it harder to induce radial displacements. Recent simulations by Braithwaite and collaborators have shown that, in stably stratified stars, random initial magnetic fields evolve into nearly axisymmetric configurations with both poloidal and toroidal components, which then remain stable for some time. It is desirable to provide an analytic study of the stability of such fields. We write an explicit expression for a plausible equilibrium structure of an axially symmetric magnetic field with both poloidal and toroidal components of adjustable strengths, in a non-barotropic, non-rotating, fluid star, and study its stability using the energy principle. We construct a displacement field that should be a reasonable approximation to the most unstable mode of a toroidal field, and confirm Braithwaite's result that a given toroidal field can be stabilized by a poloidal field containing much less energy than the former, as given through the condition $E_{\rm pol}/E_{\rm tor} \gtrsim 2 a E_{\rm tor}/E_{\rm grav}$, where $E_{\rm pol}$ and $E_{\rm tor}$ are the energies of the poloidal and toroidal fields, respectively, and $E_{\rm grav}$ is the gravitational binding energy of the star. We find that $a\approx 7.4$ for main-sequence stars, and $a \sim 200$ for neutron stars. Since $E_{\rm pol}/E_{\rm grav} \ll 1$, we conclude that the energy of the toroidal field can be substantially larger than that of the poloidal field, which is consistent with the speculation that the toroidal field is the main reservoir powering magnetar activity. The deformation of a neutron star caused by the hidden toroidal field can also cause emission of gravitational waves.
\end{abstract}


\section{Introduction}
Upper main-sequence stars, white dwarfs, and neutron stars are known to possess magnetic fields that persist for long periods of time, comparable to their lifetimes. Since convection does not play an important role in these objects, dynamo generation of magnetic fields is not expected during most of their lives. As a consequence, their magnetic fields must be in stable hydromagnetic equilibrium. However, from a theoretical perspective this poses a problem, since it can be shown that simple magnetic field configurations consisting of purely poloidal (meridional) or purely toroidal (azimuthal) fields are always unstable. In particular, Tayler (1973) showed that toroidal fields are prone to the \emph{interchange} (axisymmetric) and \emph{kink} (non-axisymmetric) instabilities. Markey \& Tayler (1973) and Wright (1973) showed similarly that purely poloidal fields, with some field lines closing inside the star, are also unstable near the \emph{neutral line}, where the poloidal field vanishes. Flowers \& Ruderman (1977) discussed another large-scale instability of poloidal fields, illustrated by the fact that, when two magnets are aligned, they will tend to orient in opposite direction to one another. Therefore, a rotation of an entire hemisphere of a star, cut along a plane containing the axis, should lead to the monotonic decrease of the overall energy, as we have demonstrated mathematically (Marchant, Reisenegger \& Akg\"{u}n 2011).

On the other hand, even in the most strongly magnetized stars, the magnetic (Lorentz) force inferred from the surface field strengths is still typically a million times weaker than the hydrostatic force due to pressure and gravity. Therefore, a small perturbation in the non-magnetic background equilibrium could be sufficient to balance the magnetic force. In the radiative envelopes of massive stars and in the interiors of degenerate stars, matter is \emph{non-barotropic} (i.e.\ pressure depends on a second quantity, such as chemical composition or specific entropy, in addition to density) and \emph{stably stratified}, allowing for a wider range of magnetic field structures than found in \emph{barotropic} fluids (i.e.\ those where pressure can be expressed as a function of density only) (Reisenegger 2009). Stable stratification also helps stabilize the magnetic field by making it harder to induce radial displacements of the fluid. This effect was included by Tayler (1973); however, by itself it is not sufficient to completely stabilize a purely toroidal (or purely poloidal) magnetic field. Recent simulations for stably stratified stars have demonstrated that initially random magnetic fields tend to evolve into nearly axisymmetric configurations with both poloidal and toroidal components of comparable strength, which then remain stable for several Alfv\'{e}n times (Braithwaite \& Spruit 2004; Braithwaite \& Nordlund 2006). In addition, Braithwaite (2009) has performed numerical tests yielding limits on the relative strengths of the two components required to stabilize each other. Our goal is to provide an analytic justification for the stability of such fields, and to understand how the poloidal and toroidal components can help stabilize each other.

This is not only relevant from the purely conceptual point of view, but has astrophysical consequences. On the one hand, it has long been speculated that magnetars contain a stronger, hidden magnetic field component that would provide the energy for their intense activity (Thompson \& Duncan 2001). This component could be the toroidal component, which, unlike its poloidal counterpart, is not visible at the surface, but whose strength should be bounded, both from above and below, by the condition of mutual stabilization. In addition, the magnetic field deforms the star, producing a mass quadrupole moment that leads to the emission of gravitational waves if the star rotates. This effect has been studied in Mastrano et al.\ (2011) for the magnetic field structures presented here.

Before considering the stability, we must first determine the equilibrium structure of the magnetic field. In \emph{barotropic} stars, the equilibrium form of the magnetic field is severely restricted and is given as the solution of a differential equation (the so-called Grad--Shafranov equation, as discussed, for example, in Chandrasekhar \& Fermi 1953; Ferraro 1954; L\"{u}st \& Schl\"{u}ter 1954; Prendergast 1956; a detailed discussion is also given in Akg\"{u}n \& Wasserman 2008). On the other hand, in realistic, \emph{non-barotropic} stars, the hydrostatic force includes a buoyancy term, which acts as a restoring force for stably stratified fluids. Thus, the only restriction that remains for an axisymmetric field in a non-barotropic fluid is that the magnetic force cannot have an azimuthal ($\hphi$) component, since no counterpart exists in the hydrostatic force that can act to balance it. In addition, the equilibrium magnetic field needs to satisfy boundary conditions at the surface and regularity conditions at the center of the star. We can construct simple polynomial forms for the scalar functions that describe the poloidal and toroidal components of the magnetic field, consistent with these requirements.

Once we know the equilibrium structure of the magnetic field, we can examine its stability, for which we use the \emph{energy principle} developed by Bernstein et al.\ (1958). In this method, one considers the energy of perturbations around the magnetic equilibrium. If this energy is always positive, then the equilibrium is stable; otherwise, it is unstable. We then consider the problem of constructing a displacement field that gives rise to instabilities in a purely toroidal field configuration. The hydrostatic and toroidal parts of the energy can be examined analytically for stability, and can be minimized with respect to the azimuthal component of the displacement field, in an analogous manner to Tayler (1973). Once we have found this minimum, we add the poloidal part of the energy and determine how strong the poloidal field must be in comparison to the toroidal field in order to stabilize the field.

The outline of this paper is as follows. In \S\ref{section_equilibrium} we discuss the equilibrium structure of the star and the magnetic field. We first construct sample equilibrium profiles for the pressure, density, and gravitational potential, which, while being sufficiently simple, have all the desirable qualities. Next, we consider the structure of the poloidal and toroidal fields and discuss their properties. We then construct a simple magnetic field that satisfies the boundary and regularity conditions. In \S\ref{section_stability} we consider the stability of the magnetic field thus constructed using the energy principle approach. We calculate the contributions to the energy due to the fluid, and due to the poloidal and toroidal components of the magnetic field. We give a proof that all physically relevant, purely toroidal fields are unstable. We discuss the implications of stable stratification on the displacement field. We construct a particular displacement field that makes the sum of the hydrostatic and toroidal parts of the energy negative, yielding an instability, and then show how the addition of a poloidal field eliminates this instability. In \S\ref{section_conclusions} we present our conclusions.

\section{Equilibrium}\label{section_equilibrium}
In realistic stars, the stress due to the magnetic field is much weaker than the hydrostatic terms due to pressure and gravity (e.g.\ Reisenegger 2009). The background equilibrium in the absence of magnetic fields and rotation is spherically symmetric and is given by Euler's equation,
        \beq
        \m\nabla P_0 + \varrho_0\m\nabla\Phi_0 = 0 \ ,
        \label{euler}
        \enq
where $P$ is pressure, $\varrho$ is density, and $\Phi$ is gravitational potential. Throughout this paper, we will denote the spherically symmetric non-magnetic background quantities with the subscript $0$. The gravitational potential is given in terms of the density by Poisson's equation,
        \beq
        \nabla^2\Phi_0 = 4\pi G\varrho_0 \ .
        \label{poisson}
        \enq

The magnetic field $\m{B}$ changes the background quantities slightly, and the new equilibrium is given by
        \beq
        \m\nabla P + \varrho\m\nabla\Phi = \frac{\m{J}\times\m{B}}{c} \ ,
        \label{equilibrium_mag}
        \enq
where $\m{J} = c \m\nabla\times\m{B}/4 \pi$ is the current density. We can express the small changes due to the magnetic field as Eulerian perturbations, and write $P = P_0 + P_1$, and similarly for $\varrho$ and $\Phi$. Then, we can rewrite the above equation, working to first order in the perturbations, as
        \beq
        \m\nabla P_1 + \varrho_1\m\nabla\Phi_0 + \varrho_0\m\nabla\Phi_1 = \frac{\m{J}\times\m{B}}{c} \ .
        \label{equilibrium_mag2}
        \enq

In the often used, idealized assumption of \emph{barotropic} fluids, there is a unique relation between pressure and density, which holds throughout the application of small perturbations. Therefore, we can write the pressure as a function of the density. This allows us to express the left-hand side of equation (\ref{equilibrium_mag}) (and consequently equation \ref{equilibrium_mag2}) as a gradient of the form $\m\nabla P + \varrho\m\nabla\Phi = \varrho\m\nabla (H + \Phi)$, where $dH(\varrho) = dP(\varrho)/\varrho$. This implies that $\m\nabla\times(\m J\times\m B/\varrho c) = 0$, so the magnetic acceleration must also be expressible as a gradient. This is a strong constraint and greatly restricts the possible choice of the magnetic field in equilibrium (Chandrasekhar \& Fermi 1953; Ferraro 1954; L\"{u}st \& Schl\"{u}ter 1954; Prendergast 1956; Akg\"{u}n \& Wasserman 2008; Haskell et al.\ 2008).

On the other hand, in \emph{non-barotropic} fluids, pressure depends on at least one additional quantity, as well as density. In white dwarfs and in the radiative zones of non-degenerate stars, the dominant additional quantity is the specific entropy, and in neutron stars it is the composition (fraction of protons or other ``impurities''; Reisenegger 2009). For long equilibration times, any changes induced in the background quantities will imply that a simple relation between pressure and density no longer exists. Consequently, the left-hand sides of equations (\ref{equilibrium_mag}) and (\ref{equilibrium_mag2}) are not expressible as gradients. Therefore, unlike the barotropic case, we do not require that the magnetic acceleration be expressible as a gradient. Instead, the only constraint for axisymmetric fields is the much less restrictive requirement that the $\hphi$ component of the magnetic force density vanish, since there is no such component in the hydrostatic part that can balance it (Chandrasekhar \& Prendergast 1956; Mestel 1956).

\subsection{Non-magnetic equilibrium}\label{section_nonmagnetic}
\begin{figure}
        \centerline{\includegraphics[scale=1.4]{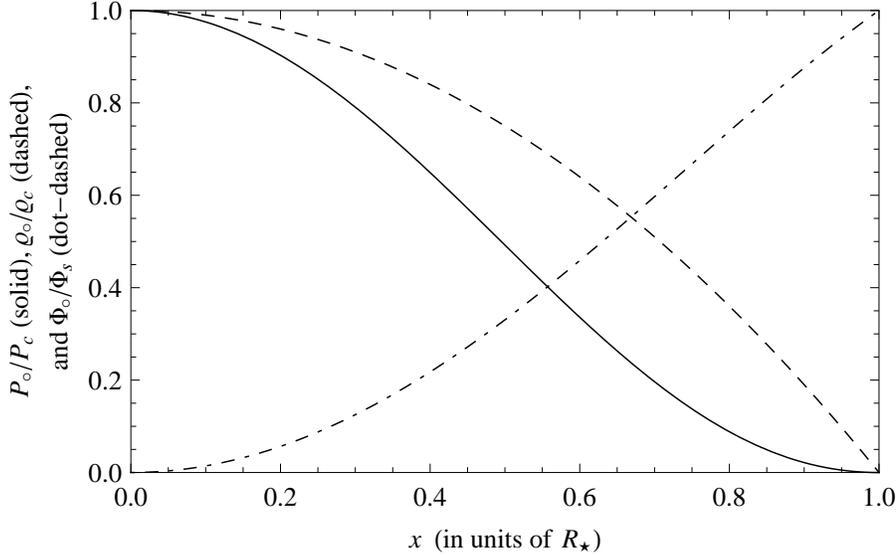}}
        \caption{Pressure, density, and gravitational potential profiles chosen for the non-magnetic equilibrium. The pressure and density are scaled by their central values, and the gravitational potential is scaled by its surface value $\Phi_s = 7P_c/4\varrho_c = 7GM_\star/8R_\star$.}
        \label{fig_pressure}
        \end{figure}

In this section, we will consider a simple model for the non-magnetic background equilibrium quantities. The derivations that follow in the subsequent sections do not rely on the specific model, but it will be needed later in the calculation of numerical estimates. We use a density profile of the form
        \beq
        \varrho_0(x) = \varrho_c(1 - x^2) \ ,
        \label{density}
        \enq
where $\varrho_c$ is the central density, and we define a dimensionless radial coordinate $x = r/R_\star$, where $R_\star$ is the stellar radius. This density profile is both simple and, at the same time, reasonably realistic, as it decreases monotonically with radius, satisfying $d\varrho_0/dx = 0$ at the center and $\varrho_0 = 0$ at the surface, and does not deviate by more than a few percent from an $n = 1$ polytrope (Mastrano et al.\ 2011). The mass enclosed within radius $x$ is given by
        \beq
        m_0(x) = 4\pi R_\star^3\int_0^x \varrho_0(x) x^2 dx = \frac{4\pi R_\star^3\varrho_c}{15} (5x^3 - 3x^5) \ .
        \enq
If the total mass of the star is denoted by $M_\star$, then the central density is $\varrho_c = 15 M_\star/8\pi R_\star^3$. The gravitational potential inside the star is given by Poisson's equation (equation \ref{poisson}),
        \beq
        \Phi_0(x) = \frac{G}{R_\star}\int_0^x \frac{m_0(x)}{x^2}dx = \frac{GM_\star}{8R_\star}(10 x^2 - 3 x^4) \ .
        \label{gravity}
        \enq
Here, the gravitational potential at the center is chosen to be zero. From Euler's equation (equation \ref{euler}), we find that the pressure is given by
        \beq
        P_0(x) = P_c - \frac{G}{R_\star}\int_0^x\frac{\varrho_0(x)m_0(x)}{x^2}dx
        = P_c\left(1 - \frac{5x^2}{2} + 2x^4 - \frac{x^6}{2}\right) \ .
        \enq
The value of the central pressure $P_c$ is determined by requiring that at the surface $P_0(1) = 0$, which yields $P_c = 4\pi G\varrho_c^2 R_\star^2/15 = 15 GM_\star^2/16\pi R_\star^4$. The profiles of the background quantities $P_0$, $\varrho_0$ and $\Phi_0$ are shown in Fig. \ref{fig_pressure}. The gravitational binding energy of the star is
        \beq
        E_{\rm grav} = 4\pi G R_\star^2 \int_0^1 x \varrho_0(x) m_0(x) dx
        = \frac{5GM_\star^2}{7R_\star} = \frac{16\pi P_c R_\star^3}{21} \ .
        \label{energy_bind}
        \enq

\subsection{Magnetic field structure}\label{section_magnetic}
The magnetic field is divergenceless, therefore quite generally it can be expressed as the sum of a \emph{poloidal} and a \emph{toroidal} component, each completely described by a single scalar function (Chandrasekhar 1981). These functions are analogous to the stream functions describing incompressible flows in hydrodynamics. An axisymmetric magnetic field can be written in spherical coordinates $(r,\theta,\phi)$ as
        \beq
        \m{B} = \m{B}_{\rm pol} + \m{B}_{\rm tor} =
        \m\nabla\alpha(r,\theta)\times\m\nabla\phi + \beta(r,\theta)\m\nabla\phi \ .
        \label{pol_tor}
        \enq
Here, we make use of the relation $\m\nabla\phi = \hphi/(r\sin\theta)$, which simplifies the calculation of curls. The current densities corresponding to each component are given by
        \beq
        \begin{split}
        \frac{4\pi\m{J}_{\rm pol}}{c} & = \m\nabla\times\m{B}_{\rm pol}
        = - \triangle\alpha\m\nabla\phi \ , \\
        \frac{4\pi\m{J}_{\rm tor}}{c} & = \m\nabla\times\m{B}_{\rm tor}
        = \m\nabla\beta\times\m\nabla\phi \ .
        \end{split}
        \label{pol_tor_current}
        \enq
We have introduced the so-called Grad--Shafranov operator, defining the cylindrical radius as $\varpi = r\sin\theta$,
        \beq
        \triangle = \varpi^2\m\nabla\cdot(\varpi^{-2}\m\nabla)
        = \pa_r^2 + \frac{\sin\theta}{r^2}\pa_\theta\left(\frac{\pa_\theta}{\sin\theta}\right) \ .
        \label{operator}
        \enq
The curl of a poloidal field is a toroidal field, and the curl of a toroidal field is a poloidal field. Therefore, $\m{J}_{\rm pol}$ is actually a toroidal field, and $\m{J}_{\rm tor}$ is a poloidal field.

We have $\m{J}_{\rm pol} \parallel \m{B}_{\rm tor} \parallel \hphi$, therefore the term $\m{J}_{\rm pol}\times\m{B}_{\rm tor}$ always vanishes in the Lorentz force. On the other hand, for the poloidal components we have $\m{J}_{\rm tor} \perp \hphi$ and $\m{B}_{\rm pol} \perp \hphi$, which implies that $\m{J}_{\rm pol}\times\m{B}_{\rm pol} \perp \hphi$ and $\m{J}_{\rm tor}\times\m{B}_{\rm tor} \perp \hphi$, while $\m{J}_{\rm tor}\times\m{B}_{\rm pol} \parallel \hphi$. However, in axisymmetric equilibrium the Lorentz force cannot have a $\hphi$ component, as implied by equation (\ref{equilibrium_mag}). Therefore, we must also have $\m{J}_{\rm tor} \parallel \m{B}_{\rm pol}$, or equivalently $\m\nabla\alpha\parallel\m\nabla\beta$, which implies that $\beta$ can be expressed as a function of $\alpha$, i.e.\ $\beta = \beta(\alpha)$. Thus, the Lorentz force can be written as the sum of a term entirely due to the poloidal field, and one entirely due to the toroidal field, $\m{f}_{\rm mag} = \m{f}_{\rm pol} + \m{f}_{\rm tor}$, where
        \beq
        \begin{split}
        4\pi\m{f}_{\rm pol} & = (\m\nabla\times\m{B}_{\rm pol})\times\m{B}_{\rm pol}
        = - \varpi^{-2}\triangle\alpha\m\nabla\alpha \ , \\
        4\pi\m{f}_{\rm tor} & = (\m\nabla\times\m{B}_{\rm tor})\times\m{B}_{\rm tor}
        = - \varpi^{-2}\beta\m\nabla\beta = - \varpi^{-2}\beta\frac{d\beta}{d\alpha}\m\nabla\alpha \ .
        \end{split}
        \label{force}
        \enq
Note that the two terms are poloidal and parallel. Moreover, they are perpendicular to the magnetic surfaces (defined as the surfaces of constant $\alpha$ and $\beta$, which contain all the field lines).

\subsection{Poloidal field}\label{section_poloidal}
In this section, we derive a simple profile for an axisymmetric poloidal magnetic field that conforms to certain boundary and regularity conditions. In particular, we impose that there are no surface currents (which would be dissipated very quickly), implying that the poloidal field is continuous across the surface. We assume that the current density drops continuously towards the surface, as the number density of charged particles should be decreasing with the mass density. Moreover, the magnetic field and current density should remain finite and continuous everywhere in the interior, and in particular at the center of the star. In what follows, we construct a particular magnetic field configuration that satisfies these requirements and appears to be at least qualitatively consistent with those found numerically by Braithwaite \& Spruit (2004) and Braithwaite \& Nordlund (2006). We warn, however, that there is a substantial arbitrariness in our choice, which we will discuss as well.

Writing the dimensional part of the magnetic field explicitly in terms of some constant $B_{\rm o}$, the poloidal field can be expressed as $\m{B}_{\rm pol} = B_{\rm o}\hat{\m\nabla}\hat{\alpha}\times\hat{\m\nabla}\phi$ (equation \ref{pol_tor}). Here, hats denote that the operators are with respect to the dimensionless radial coordinate $x = r/R_\star$, and $\hat{\alpha}$ is also dimensionless. We make our first strong assumption by taking the field outside the star to be that of a point dipole, $\m{B}_{\rm dip} \propto x^{-3}(2\hr\cos\theta + \htheta\sin\theta)$, corresponding to $\hat{\alpha}(x,\theta) \propto \sin^2\theta/x$. In order to match the angular dependence of the field on the surface, we make a second strong assumption, choosing
        \beq
        \hat{\alpha}(x,\theta) = f(x)\sin^2\theta \ .
        \label{alpha}
        \enq
The poloidal field becomes (equation \ref{pol_tor})
        \beq
        \m{B}_{\rm pol} = B_{\rm o}\hat{\m\nabla}\hat{\alpha}\times\hat{\m\nabla}\phi
        = B_{\rm o} \left[ \frac{2 f(x) \cos\theta}{x^2} \hr - \frac{f'(x) \sin\theta}{x} \htheta \right] \ .
        \label{pol_field}
        \enq
The current density is (equation \ref{pol_tor_current})
        \beq
        \frac{4\pi\m{J}_{\rm pol}}{c} = \m\nabla\times\m{B}_{\rm pol} = - \frac{B_{\rm o}}{R_\star}\hat\triangle\hat{\alpha}\hat{\m\nabla}\phi \ ,
        \label{pol_current}
        \enq
where, from equation (\ref{operator}), we have
        \beq
        \hat\triangle\hat{\alpha} = \left(f'' - \frac{2f}{x^2}\right)\sin^2\theta \ .
        \enq
Outside the surface, the current density is zero, which implies that
        \beq
        f'' = \frac{2f}{x^2} \mtext{for} x > 1 \ .
        \label{boundary1}
        \enq
Plugging in a trial solution of the form $f \propto x^s$, we find that the solutions are $s = -1$ and $s = 2$. Thus, outside the star, the solution that remains finite is given by $f \propto x^{-1}$, so, we recover the assumed point dipole. (The case $s = 2$ corresponds to a constant magnetic field in the $\hz$ direction.)

Since the density of charged particles decreases to zero at the surface of the star, there cannot be surface currents and the current density has to approach zero at the surface, implying that equation (\ref{boundary1}) must be satisfied also as $x \to 1$. In addition, the magnetic field must be continuous across the surface, which implies that both $f$ and $f'$ should be continuous. Since $f \propto x^{-1}$ outside, it then follows that
        \beq
        f' = - \frac{f}{x} \mtext{at} x = 1 \ .
        \label{boundary2}
        \enq
Since $f(1) \ne 0$, this equation requires that $|f(x)|$ decrease locally towards the surface. Moreover, we have $|f(0)| = 0 \leqslant |f(1)|$, which implies that $|f(x)|$, or equivalently $|\hat{\alpha}(x,\theta)|$, has at least one maximum somewhere within the star.

In addition to the boundary conditions at the stellar surface (equations \ref{boundary1} and \ref{boundary2}), the function $f$ must also satisfy regularity conditions at the center. Since the force density must remain finite, both the magnetic field and the current density must remain finite as well. In particular, for a trial solution of the form $f \propto x^s$, we have $\m{B}_{\rm pol} \propto x^{s-2}(2\hr\cos\theta - s\htheta\sin\theta)$ (equation \ref{pol_field}), and $4\pi\m{J}_{\rm pol}/c \propto - (s+1)(s-2)x^{s-3}\hphi\sin\theta$ (equation \ref{pol_current}). In order to avoid singularities and multi-valued functions at the origin, we must have either $s = 2$ (corresponding to the zero current case), or $s > 3$. Consistent with this, we make our third strong assumption, seeking a solution of the form
        \beq
        f(x) = f_2 x^2 + f_4 x^4 + f_6 x^6 \ .
        \enq
We need at least three terms in this polynomial ansatz, in order to be able to satisfy the two homogeneous boundary conditions at the surface (equations \ref{boundary1} and \ref{boundary2}). Considering the solution outside the star, and normalizing $f(1) = 1$, we then have
        \beq
        f(x) = \left\{\begin{array}{cl}\displaystyle\vspace{0.2cm}
        \frac{35}{8}x^2 - \frac{21}{4}x^4 + \frac{15}{8}x^6 & \ \mbox{for} \hspace{0.6cm} x \leqslant 1 \ , \\
        \displaystyle x^{-1} & \ \mbox{for} \hspace{0.6cm} x > 1 \ .
        \end{array}\right.
        \label{dipole_func}
        \enq
Somewhat more general models with additional terms are considered in the Appendix, where we also illustrate some of the different types of magnetic field configurations that can be constructed. We note that the allowed forms for $f(x)$ are entirely independent of the density profile chosen in equation (\ref{density}).

Poloidal field lines are lines of constant $\hat{\alpha}$, and are illustrated in Fig. \ref{fig_field} for the field configuration discussed here. Note that, for a given $x$, $\hat{\alpha}$ is largest along the equator. It increases smoothly from 0 at the center, reaches a maximum at $x_{\rm max} = \sqrt{(14 - \sqrt{21})/15} \approx 0.792$, where its value is $\hat{\alpha}_{\rm max} = f_{\rm max} = (931 + 21\sqrt{21})/900 \approx 1.14$, and then decreases back down to 1 at the surface. The equatorial circle of radius $x_{\rm max}$ is known as the \emph{neutral line}, and the poloidal magnetic field vanishes there.

On the other hand, the equation $\hat{\alpha}(x,\theta) = 1$ defines the last magnetic surface that closes within the star. Consequently, the region where $1 \leqslant \hat{\alpha} \leqslant \hat{\alpha}_{\rm max}$ is occupied by field lines closing inside the star. The radial extent of this region is largest along the equator, and its limits are given through the roots of $f(x) = 1$ in the interval $0 \leqslant x \leqslant 1$, which are $x = \sqrt{(27 - \sqrt{249})/30} \approx 0.612$ and $x = 1$. The largest angular extent is given by the condition $1/f_{\rm max} \leqslant \sin^2\theta$, which yields $1.21 \lesssim \theta \lesssim 1.93$ in radians (or, $69.4^\circ \lesssim \theta \lesssim 110.6^\circ$). As discussed in the Appendix, the region of closed field lines can be made larger or smaller by including more terms in $f(x)$.

\subsection{Toroidal field}\label{section_toroidal}
\begin{figure}
        \centerline{\includegraphics[scale=1]{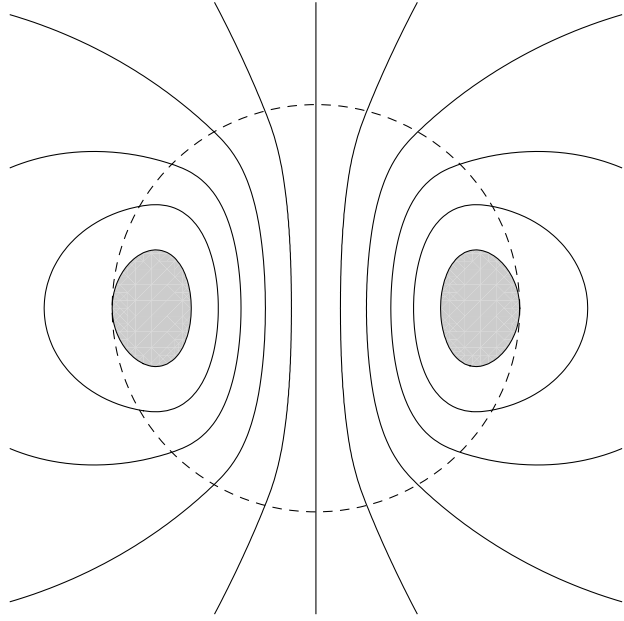}}
        \caption{Magnetic field lines for a poloidal field given through equations (\ref{alpha}) and (\ref{dipole_func}). $\alpha$ is constant along the field lines. The toroidal field is present only in the shaded donut-shaped region within the star. The stellar surface is shown with a dashed line. The field outside the star is that of a dipole (which is curl-free, i.e.\ there are no currents outside the star).}
        \label{fig_field}
        \end{figure}

An axisymmetric magnetic field with poloidal and toroidal components can be written as (equation \ref{pol_tor})
        \beq
        \m{B} = B_{\rm o}\left(\eta_{\rm pol}\hat{\m\nabla}\hat{\alpha}\times\hat{\m\nabla}\phi + \eta_{\rm tor}\hat{\beta}\hat{\m\nabla}\phi\right) \ ,
        \label{magnetic_field}
        \enq
where $\eta_{\rm pol}$ and $\eta_{\rm tor}$ are dimensionless constants that determine the relative strengths of the two components of the magnetic field. As discussed in \S\ref{section_magnetic}, $\hat{\beta}$ must be expressible as a function of $\hat{\alpha}$. Moreover, the toroidal field must vanish outside the star, since there are no currents to support it there. Since both $\hat{\alpha}$ and $\hat{\beta}$ are constant along the poloidal field lines, it follows that the toroidal field is non-zero only in a torus-shaped region defined by the poloidal field lines that close inside the star (Fig. \ref{fig_field}). The boundary conditions on the poloidal field remain unchanged, and we can still use the results of the previous section (equations \ref{alpha} and \ref{dipole_func}). In this case, the last poloidal field line that is closed within the star is given by $\hat{\alpha}(x,\theta) = \hat{\alpha}(1,\pi/2) = 1$. As a fourth strong assumption, we consider a simple relation of the form
        \beq
        \hat{\beta} = \left\{\begin{array}{cl}
        (\hat{\alpha} - 1)^n & \ \mbox{for} \hspace{0.6cm} \hat{\alpha} \geqslant 1 \ , \\
        0 & \ \mbox{for} \hspace{0.6cm} \hat{\alpha} < 1 \ .
        \end{array}\right.
        \label{beta}
        \enq
In order to avoid fast Ohmic dissipation, the current density inside the star must be continuous across the boundary where the toroidal field vanishes. This implies that we must have $n > 1$, so that the current due to the toroidal field decreases smoothly to zero at the boundary. In this paper we will consider the case $n = 2$, but whenever possible we will keep track of the power $n$ for completeness. The equilibrium pressure and density perturbations corresponding to this field structure are calculated in Mastrano et al.\ (2011).

\subsection{Amplitude of the magnetic field}
The poloidal field is largest along the axis and has a maximum at the origin, while the toroidal field is largest along the equator and has a maximum at $x \approx 0.782$. In our notation, the largest amplitudes of the two components are
        \beq
        \left(B_{\rm pol}\right)_{\rm max} = \frac{35}{4} \eta_{\rm pol} B_{\rm o} \equiv b_{\rm pol} B_{\rm o} \mtext{and}
        \left(B_{\rm tor}\right)_{\rm max} \approx 0.0254 \ \eta_{\rm tor} B_{\rm o} \equiv b_{\rm tor} B_{\rm o} \ .
        \label{amplitudes}
        \enq
The coefficients $b_{\rm pol}$ and $b_{\rm tor}$ defined in this way are dimensionless. We also note that the surface magnetic field (which is entirely poloidal) has a maximal amplitude of $2\eta_{\rm pol} B_{\rm o}$ at the poles (where it is radial), and a minimal amplitude of $\eta_{\rm pol} B_{\rm o}$ at the equator (where it is tangential to the surface).

Consider the energies stored in the poloidal and toroidal components of the magnetic field,
        \beq
        \begin{split}
        E_{\rm pol} & = \frac{1}{8\pi}\int\left|\m{B}_{\rm pol}\right|^2 dV
        = \frac{70}{33} B_{\rm o}^2 R_\star^3\eta_{\rm pol}^2
        = 2.77\times 10^{-2} B_{\rm o}^2 R_\star^3 b_{\rm pol}^2 \ , \\
        E_{\rm tor} & = \frac{1}{8\pi}\int\left|\m{B}_{\rm tor}\right|^2 dV \approx
        4.12 \times 10^{-6} B_{\rm o}^2 R_\star^3\eta_{\rm tor}^2
        \approx 6.39\times 10^{-3} B_{\rm o}^2 R_\star^3 b_{\rm tor}^2 \ .
        \end{split}
        \label{energy_mag}
        \enq
The integration for the poloidal part is carried over all of space, while the volume where the toroidal field is present is much smaller.

\section{Stability}\label{section_stability}
To study the stability of the magnetic field, consider small fluid displacements around the equilibrium given by equation (\ref{equilibrium_mag}),
        \beq
        - \varrho \frac{d^2\m\xi}{dt^2} = \varrho\omega^2\m\xi = \delta\left(\m\nabla P + \varrho\m\nabla\Phi - \m f_{\rm mag}\right) \equiv - \m{\cal F}(\m\xi) \ .
        \label{perturbations}
        \enq
Here, $\delta$ denotes Eulerian perturbations due to the displacement field $\m\xi$, and $\m{\cal F}$ is the net force density induced by the displacements. Note that there are two types of perturbations in our treatment: the magnetically induced ones with respect to the non-magnetic equilibrium, which we denote by the subscript 1 as in equation (\ref{equilibrium_mag2}), and those induced by the small displacement $\m\xi$ with respect to the magnetic equilibrium. The latter can be described either as Eulerian perturbations $\delta$ (changes at fixed locations) or Lagrangian perturbations $\Delta$ (changes as a fluid element is displaced), which are related through $\Delta = \delta + \m\xi\cdot\m\nabla$ (Friedman \& Schutz 1978).

There are two ways along which one can proceed from equation (\ref{perturbations}) in order to determine the stability of the magnetic field configuration. One method is to solve the equation for the perturbations explicitly to determine the frequencies $\omega$, and require them to be all real, $\omega^2 \geqslant 0$. Another method is to employ the energy principle of Bernstein et al.\ (1958), which has the advantage that one does not need to actually solve the equation; however, it also has the drawback that it is often quite complicated to draw general conclusions. The energy of the perturbations can be written as the sum of hydrostatic and magnetic terms, $\delta W = - \frac{1}{2}\int\m\xi\cdot\m{\cal F} d V = \delta W_{\rm hyd} + \delta W_{\rm mag}$, where (Akg\"{u}n \& Wasserman 2008)
        \beq
        \begin{split}
        \delta W_{\rm hyd} = & \frac{1}{2}\int\left[\Gamma P (\m\nabla\cdot\m\xi)^2
        + (\m\xi\cdot\m\nabla P)(\m\nabla\cdot\m\xi)
        - (\m\xi\cdot\m\nabla\Phi)(\m\nabla\cdot\varrho\m\xi)
        + \varrho\m\xi\cdot\m\nabla\delta\Phi\right] d V \\
        & - \frac{1}{2}\oint\left[\Gamma P\m\nabla\cdot\m\xi + \m\xi\cdot\m\nabla P\right] \m\xi\cdot d\m{S} \ , \\
        \delta W_{\rm mag}
        = & \frac{1}{2}\int\left[\frac{|\delta\m{B}|^2}{4\pi} -
        \frac{\m{J}\cdot\delta\m{B}\times\m\xi}{c}\right]d V
        + \frac{1}{8\pi}\oint [\m\xi(\m{B}\cdot\delta\m{B})
        - \m{B}(\m\xi\cdot\delta\m{B})]\cdot d\m{S} \ .
        \end{split}
        \label{energy_pert}
        \enq
The magnetic field perturbation follows from Faraday's law of induction,
        \beq
        \delta\m{B} = \m\nabla\times(\m\xi\times\m{B}) \ .
        \label{Faraday}
        \enq

\subsection{Implications of non-barotropy}
The pressure in a non-barotropic fluid can be written as $P(\varrho,s)$, where $s$ is the specific entropy or chemical composition, depending on the type of star (as discussed in \S\ref{section_equilibrium}). In the non-magnetic background equilibrium, this quantity is a function of density, $s_0(\varrho_0)$, because both $s_0$ and $\varrho_0$ are functions of radius. Thus, the background equilibrium is described by a single index,
        \beq
        \gamma = \frac{d\ln P_0}{d\ln\varrho_0} = \left(\frac{\pa\ln P}{\pa\ln\varrho}\right)_s
        + \left(\frac{\pa\ln P}{\pa\ln s}\right)_\varrho \frac{d\ln s_0}{d\ln\varrho_0} \ .
        \label{gamma}
        \enq
For the non-magnetic equilibrium described in \S\ref{section_nonmagnetic}, we have $\gamma(x) = (5 - 3x^2)/(2 - x^2)$, which decreases monotonically from $\gamma(0) = 5/2$ to $\gamma(1) = 2$.

For long equilibration times, the quantity $s$ of a given fluid element remains constant as it is displaced, therefore $\Delta s = 0$. Then, the Lagrangian perturbations of pressure and density are related through
        \beq
        \frac{\Delta P}{P} = \left(\frac{\pa\ln P}{\pa\ln\varrho}\right)_s\frac{\Delta\varrho}{\varrho} \equiv \Gamma\frac{\Delta\varrho}{\varrho} \ .
        \label{gamma_ad}
        \enq
Similarly, working to lowest order in $B^2$ (dropping terms of the order $\xi B^2$), and using $\Delta\varrho = - \varrho\m\nabla\cdot\m\xi$, $\delta\varrho = - \m\nabla\cdot(\varrho\m\xi)$, $\delta s = - \m\xi\cdot\m\nabla s \approx - (d s_0 / d \varrho_0) \m\xi\cdot\m\nabla\varrho_0$, and the definitions of $\gamma$ and $\Gamma$, the Eulerian perturbation of pressure can be written as
        \beq
        \frac{\delta P}{P} = \left(\frac{\pa\ln P}{\pa\ln\varrho}\right)_s\frac{\delta\varrho}{\varrho}
        + \left(\frac{\pa\ln P}{\pa\ln s}\right)_\varrho\frac{\delta s}{s}
        \approx \gamma\frac{\delta\varrho}{\varrho} + (\Gamma - \gamma) \frac{\Delta\varrho}{\varrho} \ .
        \label{delta_P}
        \enq
In a non-barotropic fluid, $\Gamma \ne \gamma$, and the hydrostatic force (which we define as the sum of pressure and gravitational forces, $\m{f}_{\rm hyd} = - \m\nabla P - \varrho\m\nabla\Phi$) now gives rise to an additional term due to buoyancy, which is proportional to the difference between the indices. Upon the application of small perturbations we have
        \beq
        \delta\m{f}_{\rm hyd} = - \m\nabla\delta P - \delta\varrho\m\nabla\Phi - \varrho\m\nabla\delta\Phi
        = - \varrho\m\nabla\left(\frac{\delta P}{\varrho} + \delta\Phi\right)
        + \left(\frac{\Gamma}{\gamma} - 1\right)\Delta\varrho\m\nabla\Phi \ .
        \label{delta_f_hyd}
        \enq
In a stably stratified star $\Gamma > \gamma$, and the second term acts as a restoring force. Typically, in upper main-sequence stars $\Gamma/\gamma - 1 \sim 1/4$, in white dwarfs $\Gamma/\gamma - 1 \sim T_7/500$, where $T_7$ is the internal temperature in units of $10^7$ K, and in neutron stars $\Gamma/\gamma - 1 \sim$ few $\%$ (Reisenegger 2009 and references therein).

\subsection{Implications of stable stratification}\label{section_stratification}
Consider the integrands of the hydrostatic and magnetic parts given by equation (\ref{energy_pert}). For simplicity, we will always be concerned with cases where the surface integrals vanish (i.e.\ $\m\xi = 0$ at the surface), and we will employ the \emph{Cowling approximation} of neglecting perturbations of the gravitational potential ($\delta\Phi = 0$). Formally, both of these simplifications make us overestimate the stability of the star, which was shown for the latter by Tayler (1973) and is obvious for the former as it makes us disregard potentially unstable displacement fields affecting the stellar surface. However, we will argue that the most unstable (and thus most relevant) displacement fields are fairly localized inside the star and non-radial, so they will have very little effect on the stellar surface or the gravitational potential. We have
        \beq
        \begin{split}
        {\cal E}_{\rm hyd} & = \Gamma P (\m\nabla\cdot\m\xi)^2 + (\m\xi\cdot\m\nabla P)(\m\nabla\cdot\m\xi)
        - (\m\xi\cdot\m\nabla\Phi)(\m\nabla\cdot\varrho\m\xi) \ , \\
        {\cal E}_{\rm mag} & = \frac{1}{4\pi}\bra{[}|\delta\m{B}|^2
        - \m\xi\times(\m\nabla\times\m{B})\cdot\delta\m{B}\ket{]} \ .
        \end{split}
        \label{hydrostatic}
        \enq

We can write the adiabatic index of the perturbations as $\Gamma = \Gamma_0 + \Gamma_1$, where, from equation (\ref{gamma_ad}),
        \beq
        \Gamma = \left.\left(\frac{\pa\ln P}{\pa\ln\varrho}\right)_s \right|_{\varrho,s} \mtext{and}
        \Gamma_0 = \left.\left(\frac{\pa\ln P}{\pa\ln\varrho}\right)_s \right|_{\varrho_0,s_0} \ .
        \enq
Note that $|\Gamma_1| / \Gamma_0 \sim |P_1| / P_0 \sim |\varrho_1| / \varrho_0 \sim |\Phi_1| / \Phi_0 \sim B^2 / P_0 \lesssim 10^{-6}$ (Reisenegger 2009). Thus, the hydrostatic integrand can be rewritten as
        \beq
        \begin{split}
        {\cal E}_{\rm hyd} = & (\Gamma_0 - \gamma) P_0(\m\nabla\cdot\m\xi)^2
        + \frac{\gamma P_0}{\varrho_0^2}(\m\nabla\cdot\varrho_0\m\xi)^2 \\
        & + (\Gamma_1 P_0 + \Gamma_0 P_1) (\m\nabla\cdot\m\xi)^2
        + (\m\xi\cdot\m\nabla P_1)(\m\nabla\cdot\m\xi)
        - (\m\xi\cdot\m\nabla\Phi_0)(\m\nabla\cdot\varrho_1\m\xi)
        - (\m\xi\cdot\m\nabla\Phi_1)(\m\nabla\cdot\varrho_0\m\xi) \ .
        \end{split}
        \label{hydrostatic2}
        \enq
For stably stratified stars, $\Gamma_0 > \gamma$, so that the first two terms of the integrand are positive definite. The remaining terms of the integrand are corrections due to the magnetic field. These, as well as ${\cal E}_{\rm mag}$, can be positive or negative, but their magnitude is $\lesssim\xi^2 B^2/L^2$, where $L$ is some length scale characterizing the spatial variations of the magnetic field. In order for the total energy to be negative, thus allowing for the existence of instabilities, the first two terms of ${\cal E}_{\rm hyd}$ must also be small,
        \beq
        (\Gamma_0 - \gamma) P_0 (\m\nabla\cdot\m\xi)^2 \lesssim \frac{\xi^2 B^2}{L^2} \mtext{and}
        \frac{\gamma P_0}{\varrho_0^2} (\m\nabla\cdot\varrho_0\m\xi)^2
        \lesssim \frac{\xi^2 B^2}{L^2} \ .
        \label{div_xi}
        \enq
These are constraints that need to be satisfied by the displacement field in order to potentially lead to instabilities. They also imply the following bounds for the remaining terms in ${\cal E}_{\rm hyd}$,
        \beq
        \begin{split}
        |\Gamma_1| P_0 (\m\nabla\cdot\m\xi)^2 \sim \Gamma_0 |P_1| (\m\nabla\cdot\m\xi)^2 &
        \lesssim \frac{\Gamma_0 \xi^2 B^4}{(\Gamma_0 - \gamma)P_0 L^2} \ , \\
        |(\m\xi\cdot\m\nabla P_1)(\m\nabla\cdot\m\xi)|
        \sim |(\m\xi\cdot\m\nabla\Phi_0)(\m\nabla\cdot\varrho_1\m\xi)|
        & \lesssim \frac{\xi^2 B^3}{\sqrt{(\Gamma_0 - \gamma) P_0} L^2} \ , \\
        |(\m\xi\cdot\m\nabla\Phi_1)(\m\nabla\cdot\varrho_0\m\xi)|
        & \lesssim \frac{\xi^2 B^3}{\sqrt{\gamma P_0} L^2} \ .
        \end{split}
        \enq
Here, we assume that both $\gamma$ and $\Gamma_0$ are of order unity. Although $\Gamma_0 - \gamma \sim 10^{-2} \ll 1$ in some realistic cases, it is still much larger than the ratio of magnetic pressure to background pressure, $B^2/P_0 \sim 10^{-6}$. Thus, we conclude that (i) corrections to the equilibrium pressure and density due to the magnetic field give rise to terms in the hydrostatic energy that are at least a factor of $B/\sqrt{(\Gamma_0 - \gamma)P_0} \lesssim 10^{-2}$ smaller than the (potentially destabilizing) magnetic energy contributions, and therefore can be left out; (ii) the conditions given by equation (\ref{div_xi}) also imply that the radial component of the displacement field is small, $\xi_r^2/\xi^2 \lesssim B^2/(\Gamma_0 - \gamma)P_0 \ll 1$.

\subsection{Energy of perturbations for a general displacement field}
In this section, we will write down the energy of arbitrary perturbations for poloidal and toroidal fields. Since we assume axisymmetry, and none of the equilibrium quantities depends on the azimuthal angle $\phi$, we can express the displacement field in general as a superposition of components of the form
        \beq
        \m\xi = \bra{[}R(r,\theta)\hr + S(r,\theta)\htheta + iT(r,\theta)\hphi\ket{]}r\sin\theta e^{im\phi} \ ,
        \label{xi}
        \enq
which can be analyzed separately for different $m$ (as they do not mix in the energy). For notational convenience, we have explicitly written out a factor of cylindrical radius. In general, the dimensionless functions $R$, $S$ and $T$ will be complex, but only the real part of $\m\xi$ is physically relevant. Therefore, products should be treated as $ZZ^*$, where $^*$ denotes the complex conjugate.\footnote{Caution must be taken in using complex notation to describe real physical quantities. Here, we are dealing with functions of the form $f = F(r,\theta)e^{im\phi}$ and $g = G(r,\theta)e^{im\phi}$, and are interested in integrals of the products of their real parts (denoted by $\Re$), which can be written as
        \beq
        \int_0^{2\pi} \Re(f)\Re(g)d\phi = \frac{1}{2}\int_0^{2\pi} \Re(f g^*) d\phi = \pi\Re(F G^*) \ .
        \nonumber
        \enq
\label{footnote_cmplx}}

The energy of the perturbations can be calculated from equation (\ref{hydrostatic}). For notational convenience, define an operator $\Lambda$ and an auxiliary quantity $D_m$ by
        \beq
        \Lambda(u) = R\pa_r u + \frac{S\pa_\theta u}{r} \mtext{and}
        D_m = \frac{\pa_r(r^3 R)}{r^3} + \frac{\pa_\theta(S\sin^2\theta)}{r\sin^2\theta} - \frac{mT}{r\sin\theta} \ .
        \label{Lambda_D}
        \enq
$\Lambda$ is the directional derivative along the displacement field, $\m\xi\cdot\m\nabla u = \Lambda(u)r\sin\theta e^{im\phi}$, where $u$ is an equilibrium quantity independent of the angle $\phi$. $D_m$ is the divergence of the displacement field, $\m\nabla\cdot\m\xi = D_m r\sin\theta e^{im\phi}$, and we will explicitly keep track of its dependence on $m$. Defining $\varpi = r\sin\theta$, the hydrostatic part of the integrand (equation \ref{hydrostatic}) becomes
        \beq
        {\cal E}_{\rm hyd} = \frac{1}{2}\varpi^2 \Re
        \left\{\Gamma P D_m {D_m}^* + [\Lambda(P) - \varrho\Lambda(\Phi)]{D_m}^* - \Lambda(\varrho)\Lambda^*(\Phi)\right\} \ .
        \label{calE_hyd}
        \enq
The factor $1/2$ arises as a consequence of the complex notation, as discussed in footnote \ref{footnote_cmplx}. The equation of hydrostatic equilibrium (equation \ref{equilibrium_mag}) relates the pressure, density, and gravitational potential to the magnetic field. Therefore, ${\cal E}_{\rm hyd}$ depends implicitly on the functions $\alpha$ and $\beta$ for the poloidal and toroidal components of the magnetic field through the small corrections that these induce on the background quantities. However, as discussed in \S\ref{section_stratification}, these corrections can be dropped in the calculation of ${\cal E}_{\rm hyd}$, so the equilibrium quantities $P$, $\varrho$, and $\Phi$ in equation (\ref{calE_hyd}) can be taken as their non-magnetic versions $P_0$, $\varrho_0$, and $\Phi_0$.

The calculation of the magnetic part of the integrand is more involved. The perturbations of the poloidal and toroidal components of the magnetic field (equation \ref{pol_tor}), are given by equation (\ref{Faraday}) as
        \beq
        \begin{split}
        \delta\m{B}_{\rm pol} = & \m\nabla\times(\m\xi\times\m{B}_{\rm pol})
        = \m\nabla(\m\xi\cdot\m\nabla\phi)\times\m\nabla\alpha
        - \m\nabla(\m\xi\cdot\m\nabla\alpha)\times\m\nabla\phi \\
        = & \left\{\frac{mT\pa_\theta\alpha - \pa_\theta[\varpi\Lambda(\alpha)]}{r\varpi}\hr
        - \frac{mT\pa_r\alpha - \pa_r[\varpi\Lambda(\alpha)]}{\varpi}\htheta
        + \frac{i(\pa_r T\pa_\theta\alpha - \pa_\theta T\pa_r\alpha)}{r}\hphi\right\}e^{im\phi} \ , \\
        \delta\m{B}_{\rm tor} = & \m\nabla\times(\m\xi\times\m{B}_{\rm tor})
        = (\beta\m\nabla\phi\cdot\m\nabla)\m\xi - (\beta\m\nabla\phi)(\m\nabla\cdot\m\xi)
        - (\m\xi\cdot\m\nabla)(\beta\m\nabla\phi) \\
        = & \left\{\frac{imR\beta}{\varpi}\hr + \frac{imS\beta}{\varpi}\htheta
        - \frac{\pa_r(rR\beta) + \pa_\theta(S\beta)}{r}\hphi\right\}e^{im\phi} \ .
        \end{split}
        \enq
Also, using equation (\ref{pol_tor_current}), we have
        \beq
        \begin{split}
        \m\xi\times(\m\nabla\times\m{B}_{\rm pol}) & = \m\xi\times(- \triangle\alpha\m\nabla\phi)
        = - \triangle\alpha(S\hr - R\htheta) e^{im\phi} \ , \\
        \m\xi\times(\m\nabla\times\m{B}_{\rm tor}) & = \m\xi\times(\m\nabla\beta\times\m\nabla\phi)
        = (\m\xi\cdot\m\nabla\phi)\m\nabla\beta - (\m\xi\cdot\m\nabla\beta)\m\nabla\phi
        = [iT\m\nabla\beta - \Lambda(\beta)\hphi]e^{im\phi} \ .
        \end{split}
        \enq
The magnetic part of the integrand (equation \ref{hydrostatic}) can be written as a sum of three terms: one that is entirely due to the poloidal field, one entirely due to the toroidal field, and a third term that is a combination of the two components, ${\cal E}_{\rm mag} = {\cal E}_{\rm pol} + {\cal E}_{\rm tor} + {\cal E}_{\rm cross}$, where
        \beq
        \begin{split}
        {\cal E}_{\rm pol} & = \frac{1}{8\pi}\Re\left[\delta\m{B}_{\rm pol}^{}\cdot\delta\m{B}_{\rm pol}^*
        - \m\xi\times(\m\nabla\times\m{B}_{\rm pol}^{})\cdot\delta\m{B}_{\rm pol}^*\right] \ , \\
        {\cal E}_{\rm tor} & = \frac{1}{8\pi}\Re\left[\delta\m{B}_{\rm tor}^{}\cdot\delta\m{B}_{\rm tor}^*
        - \m\xi\times(\m\nabla\times\m{B}_{\rm tor}^{})\cdot\delta\m{B}_{\rm tor}^*\right] \ , \\
        {\cal E}_{\rm cross} & = \frac{1}{8\pi}\Re\left[\delta\m{B}_{\rm pol}^{}\cdot\delta\m{B}_{\rm tor}^*
        + \delta\m{B}_{\rm tor}^{}\cdot\delta\m{B}_{\rm pol}^*
        - \m\xi\times(\m\nabla\times\m{B}_{\rm pol}^{})\cdot\delta\m{B}_{\rm tor}^*
        - \m\xi\times(\m\nabla\times\m{B}_{\rm tor}^{})\cdot\delta\m{B}_{\rm pol}^*\right] \ .
        \end{split}
        \enq
After some algebra, we obtain
        \beq
        \begin{split}
        {\cal E}_{\rm pol} = & \frac{1}{8\pi}\left\{
        \left|\frac{mT\pa_r\alpha - \pa_r[\varpi\Lambda(\alpha)]}{\varpi} + \frac{R\triangle\alpha}{2}\right|^2
        + \left|\frac{mT\pa_\theta\alpha - \pa_\theta[\varpi\Lambda(\alpha)]}{r\varpi}
        + \frac{S\triangle\alpha}{2}\right|^2 \right. \\
        & \left. + \left|\frac{\pa_r T\pa_\theta\alpha - \pa_\theta T\pa_r\alpha}{r}\right|^2
        - \frac{(|R|^2 + |S|^2)(\triangle\alpha)^2}{4} \right\} \ , \\
        {\cal E}_{\rm tor} = & \frac{1}{8\pi}\left\{
        \left|\frac{\beta[\pa_r(r R) + \pa_\theta S]}{r} + \frac{\Lambda(\beta)}{2}\right|^2
        - \left|\frac{m\beta T}{\varpi} + \frac{\Lambda(\beta)}{2}\right|^2
        + \frac{m^2\beta^2(|R|^2 + |S|^2 + |T|^2)}{\varpi^2} \right\} \ , \\
        {\cal E}_{\rm cross} = & \frac{1}{8\pi}\Re\left\{ \frac{iT}{r\varpi}
        \bra{[}\pa_r[\varpi\Lambda(\alpha)]\pa_\theta\beta
        - \pa_\theta[\varpi\Lambda(\alpha)]\pa_r\beta\ket{]}^*
        + \frac{2i\beta}{r^2}\bra{[}\pa_r(rR) +\pa_\theta S\ket{]}
        \bra{[}\pa_r T\pa_\theta\alpha - \pa_\theta T\pa_r\alpha\ket{]}^* \right. \\
        & \left. + \frac{2im\beta}{\varpi}\left[RS^*\triangle\alpha
        + \frac{mT\pa_r\alpha - \pa_r[\varpi\Lambda(\alpha)]}{\varpi}S^*
        - \frac{mT\pa_\theta\alpha - \pa_\theta[\varpi\Lambda(\alpha)]}{r\varpi}R^*\right] \right\} \ .
        \end{split}
        \label{calE_mag}
        \enq

\subsection{Stability of a toroidal field}
In this section, following the derivation of Tayler (1973), we consider the problem of constructing a displacement field that makes a purely toroidal magnetic field unstable. In other words, we want to find $\m\xi$ for which ${\cal E}_{\rm hyd} + {\cal E}_{\rm tor} < 0$. Then, in the following section we will examine the stability of the poloidal part for the same displacement.

As demonstrated by Tayler (1973) for the purely toroidal field, the real and imaginary parts in the energy separate into two equivalent terms. Consequently, it is sufficient to consider the case of real $R$, $S$, and $T$. The function $T$ appears only algebraically in the hydrostatic and toroidal parts of the integrand. We then have, from equations (\ref{calE_hyd}) and (\ref{calE_mag}),
        \beq
        {\cal E}_{\rm hyd} + {\cal E}_{\rm tor} = \frac{1}{2}\left[E_2 (mT)^2 + E_1 mT + E_0\right] \ ,
        \label{integrand_hyd_tor}
        \enq
where we define,
        \beq
        \begin{split}
        E_2 = & \Gamma P \ , \\
        E_1 = & - 2\varpi\Gamma P D_0 - \varpi\Lambda(P) + \varpi\varrho\Lambda(\Phi)
        - \frac{\beta\Lambda(\beta)}{4\pi\varpi} \ , \\
        E_0 = & \left[\varpi^2\Gamma P + \frac{\beta^2}{4\pi}\right] D_0^2
        + \left[\varpi^2\Lambda(P) - \varpi^2\varrho\Lambda(\Phi) + \frac{\beta\Lambda(\beta)}{4\pi}
        - \frac{\beta^2\Lambda(\varpi)}{\pi\varpi}\right] D_0 \\
        & - \varpi^2\Lambda(\varrho)\Lambda(\Phi)
        - \frac{\beta\Lambda(\beta)\Lambda(\varpi)}{2\pi\varpi}
        + \frac{\beta^2\Lambda^2(\varpi)}{\pi\varpi^2}
        + \frac{m^2\beta^2(R^2 + S^2)}{4\pi\varpi^2} \ .
        \end{split}
        \label{def_E}
        \enq
Here, $D_0 = D_m + mT/r\sin\theta$ (equation \ref{Lambda_D}) is the only term that contains derivatives of the functions $R$ and $S$. The above terms can be somewhat simplified using the equation of equilibrium for purely toroidal fields, which follows from equations (\ref{equilibrium_mag}), (\ref{force}), and (\ref{Lambda_D}) as
        \beq
        \Lambda(P) + \varrho\Lambda(\Phi) = - \frac{\beta\Lambda(\beta)}{4\pi\varpi^2} \ .
        \label{Lambda_eq_tor}
        \enq

Since $E_2 > 0$, the integrand given by equation (\ref{integrand_hyd_tor}) can be minimized with respect to $T$ for $m \ne 0$. In the minimization, we hold $R$ and $S$ (and therefore $D_0$) constant. The minimizing value is $mT/\varpi = - E_1 / 2\varpi E_2 = D_0 - \varrho \Lambda(\Phi)/\Gamma P$ and the minimum of the integrand is ${\cal E}_{\rm hyd} + {\cal E}_{\rm tor} = E_0 / 2 - E_1^2 / 8E_2$. Using equation (\ref{Lambda_D}), this minimization corresponds to setting $D_m = \varrho \Lambda(\Phi)/\Gamma P$, which can be alternatively expressed as
        \beq
        \Gamma P \m\nabla\cdot\m\xi = \varrho\m\xi\cdot\m\nabla\Phi \ .
        \label{minimum_T}
        \enq
Dropping magnetic corrections to the background quantities (which give rise to terms of the order $\xi B^2$), using $\delta\varrho = - \m\nabla\cdot(\varrho\m\xi)$, $\Delta\varrho = - \varrho\m\nabla\cdot\m\xi$, and equation (\ref{delta_P}), this can be rewritten as
        \beq
        \frac{\delta P}{P} \approx \gamma\frac{\delta\varrho}{\varrho} + (\Gamma - \gamma)\frac{\Delta\varrho}{\varrho} \approx 0 \ .
        \label{minimum_T2}
        \enq
Note that ${\cal E}_{\rm hyd}$ is a quadratic function of $T$ (equation \ref{calE_hyd}) and ${\cal E}_{\rm tor}$ is a linear function of $T$ (equation \ref{calE_mag}). This implies that both ${\cal E}_{\rm hyd}$ and ${\cal E}_{\rm hyd} + {\cal E}_{\rm tor}$ can be minimized with respect to $T$ for $m \ne 0$. In fact, the minima of the non-magnetic case (which corresponds to minimizing ${\cal E}_{\rm hyd}$) and the purely toroidal case (which corresponds to minimizing ${\cal E}_{\rm hyd} + {\cal E}_{\rm tor}$) are both obtained for the condition given by equation (\ref{minimum_T}). These minima are not precisely identical since the background quantities differ by a small amount between the two cases. Equation (\ref{minimum_T2}) implies that the minimum is obtained by setting $\delta P = 0$, which in a barotropic fluid ($\Gamma = \gamma$) further implies that $\delta\varrho = 0$. The minimum of ${\cal E}_{\rm hyd}$ to lowest order is ${\cal E}_{\rm hyd} = (1/\gamma - 1/\Gamma) (\m\xi\cdot\m\nabla P_0)^2 / P_0$ (equation \ref{hydrostatic}), which is zero for a barotropic fluid, while for a stably stratified non-barotropic fluid it is positive (as long as $R \ne 0$).

On the other hand, for $m = 0$, we have ${\cal E}_{\rm hyd} + {\cal E}_{\rm tor} = E_0/2$. Thus, in general, we can combine the two cases ($m = 0$ and $m \ne 0$) and write the energy for any $m$ as
        \beq
        {\cal E}_{\rm hyd} + {\cal E}_{\rm tor} = \frac{E_0}{2} - (1 - \delta_{m0}) \frac{E_1^2}{8E_2} \ , \mtext{where}
        \delta_{m0} = \left\{\begin{array}{ll}
        1 & \ \mbox{for} \hspace{0.6cm} m = 0 \ , \\
        0 & \ \mbox{for} \hspace{0.6cm} m \ne 0 \ .
        \end{array}\right.
        \enq
We can further rewrite the integrand by grouping the $D_0$ terms together and writing them as a complete square, thus separating the derivatives of $R$ and $S$ and leaving out only algebraic terms. Defining $K_m = \delta_{m0}\Gamma P + \beta^2/4\pi\varpi^2$, we have
        \beq
        {\cal E}_{\rm hyd} + {\cal E}_{\rm tor} = \frac{1}{2}\varpi^2 K_m
        \left\{D_0 - \frac{1}{K_m}\left[\delta_{m0}\varrho\Lambda(\Phi) + \frac{\beta^2\Lambda(\varpi)}{2\pi\varpi^3}\right]\right\}^2
        + \frac{1}{2}\varpi^2(a_m R^2 + b_m R S + c_m S^2) \ .
        \label{integrand_hyd_tor_m}
        \enq
Keep in mind that this integrand is already minimized with respect to $T$ for $m \ne 0$. The first term is always positive, and the second term forms a quadratic in $R$ and $S$. The positive definite term can always be made to vanish by a suitable choice of the displacement field. Therefore, the integrand is always positive if the quadratic is positive, which corresponds to the conditions
        \beq
        a_m > 0 \ , \hspace{0.6cm} c_m > 0 \mtext{and} b_m^2 < 4 a_m c_m \ .
        \label{conditions}
        \enq
These are sufficient and necessary conditions for the \emph{stability} of the toroidal field (Tayler 1973). Note that they are not independent: one of the first two, together with the last one, imply the remaining condition. The coefficients for any $m$ are given through
        \beq
        \begin{split}
        a_m = & - \pa_r\varrho\pa_r\Phi - (1 - \delta_{m0}) \frac{\varrho^2(\pa_r\Phi)^2}{\Gamma P}
        - \frac{1}{K_m}\left( \delta_{m0}\varrho\pa_r\Phi + \frac{\beta^2}{2\pi r^3\sin^2\theta} \right)^2 \\
        & - \frac{\beta\pa_r\beta}{2\pi r^3\sin^2\theta} + \frac{\beta^2}{\pi r^4\sin^2\theta}
        + \frac{m^2\beta^2}{4\pi r^4\sin^4\theta} \ , \\
        b_m = & - \frac{\pa_r\varrho\pa_\theta\Phi}{r} - \frac{\pa_\theta\varrho\pa_r\Phi}{r}
        - (1 - \delta_{m0})\frac{2\varrho^2\pa_r\Phi\pa_\theta\Phi}{r\Gamma P} \\
        & - \frac{2}{r K_m} \left(\delta_{m0}\varrho\pa_r\Phi + \frac{\beta^2}{2\pi r^3\sin^2\theta}\right)
        \left(\delta_{m0}\varrho\pa_\theta\Phi + \frac{\beta^2\cos\theta}{2\pi r^2\sin^3\theta}\right) \\
        & - \frac{\beta\pa_r\beta \cos\theta}{2\pi r^3\sin^3\theta}
        - \frac{\beta\pa_\theta\beta}{2\pi r^4\sin^2\theta}
        + \frac{2\beta^2\cos\theta}{\pi r^4\sin^3\theta} \ , \\
        c_m = & - \frac{\pa_\theta\varrho\pa_\theta\Phi}{r^2} - (1 - \delta_{m0}) \frac{\varrho^2(\pa_\theta\Phi)^2}{r^2 \Gamma P}
        - \frac{1}{r^2 K_m} \left(\delta_{m0}\varrho\pa_\theta\Phi + \frac{\beta^2\cos\theta}{2\pi r^2\sin^3\theta}\right)^2 \\
        & - \frac{\beta\pa_\theta\beta \cos\theta}{2\pi r^4\sin^3\theta} + \frac{\beta^2\cos^2\theta}{\pi r^4\sin^4\theta}
        + \frac{m^2\beta^2}{4\pi r^4\sin^4\theta} \ .
        \end{split}
        \label{coefficients_m}
        \enq
These are equivalent to the results given by Tayler (1973), Goossens \& Veugelen (1978), and Akg\"{u}n \& Wasserman (2008), albeit the notation is somewhat different. (Here, we have combined the cases $m = 0$ and $m \ne 0$ into a single general form.) We have $|\varrho\pa_r\Phi| \sim |\Phi\pa_r\varrho| \sim P_0/R_\star$ and $|\varrho\pa_\theta\Phi| \sim |\Phi\pa_\theta\varrho| \sim B^2$, so that, to leading order, the coefficients are
        \beq
        a_m \approx \left(\frac{1}{\gamma} - \frac{1}{\Gamma}\right) \frac{(\pa_r P_0)^2}{P_0}
        \equiv \varrho_0 N^2 \sim \frac{(\Gamma - \gamma)P_0}{R_\star^2}
        \mtext{and}
        |b_m| \ , \ |c_m| \sim \frac{B^2}{4\pi R_\star^2} \equiv \varrho_0 \omega_A^2 \ .
        \label{leading_am_cm}
        \enq
Here, $N$ is the Brunt-V\"{a}is\"{a}l\"{a} frequency, and $\omega_A$ is the Alfv\'{e}n frequency (i.e.\ the inverse of the Alfv\'{e}n crossing-time for the star). In a stably stratified star, $\Gamma > \gamma$, where the two gammas, defined by equations (\ref{gamma}) and (\ref{gamma_ad}), are of order unity. In this case, the condition $a_m > 0$ is comfortably satisfied, and the problem reduces to showing whether $b_m^2 < 4 a_m c_m$ is satisfied, since, if it is true, then the remaining condition $c_m > 0$ follows trivially. However, note that, when $c_m < 0$, the magnetic field is always unstable, immaterial of the value of $b_m$. The field can also be unstable when $c_m$ is positive, but sufficiently close to zero, while $b_m$ is sufficiently large ($0 < c_m < b_m^2/4a_m \sim B^4/64\pi^2 (\Gamma - \gamma) P_0R_\star^2)$. This is a very narrow interval. For larger, positive $c_m$, the condition $b_m^2 < 4 a_m c_m$ will always be satisfied.

\subsubsection{Coefficients for $m = 0$}
For future reference, we quote the coefficients for $m = 0$ here. In this case, we have $K_0 = \Gamma P + \beta^2/4\pi\varpi^2$, and the coefficients given by equation (\ref{coefficients_m}) reduce to (Tayler 1973; Goossens \& Veugelen 1978; Akg\"{u}n \& Wasserman 2008)
        \beq
        \begin{split}
        a_0 = & - \pa_r\varrho\pa_r\Phi - \frac{1}{K_0}\left( \varrho\pa_r\Phi + \frac{\beta^2}{2\pi r^3\sin^2\theta} \right)^2
        - \frac{\beta\pa_r\beta}{2\pi r^3\sin^2\theta} + \frac{\beta^2}{\pi r^4\sin^2\theta} \ , \\
        b_0 = & - \frac{\pa_r\varrho\pa_\theta\Phi}{r} - \frac{\pa_\theta\varrho\pa_r\Phi}{r}
        - \frac{2}{r K_0} \left(\varrho\pa_r\Phi + \frac{\beta^2}{2\pi r^3\sin^2\theta}\right)
        \left(\varrho\pa_\theta\Phi + \frac{\beta^2\cos\theta}{2\pi r^2\sin^3\theta}\right) \\
        & - \frac{\beta\pa_r\beta \cos\theta}{2\pi r^3\sin^3\theta}
        - \frac{\beta\pa_\theta\beta}{2\pi r^4\sin^2\theta}
        + \frac{2\beta^2\cos\theta}{\pi r^4\sin^3\theta} \ , \\
        c_0 = & - \frac{\pa_\theta\varrho\pa_\theta\Phi}{r^2}
        - \frac{1}{r^2 K_0} \left(\varrho\pa_\theta\Phi + \frac{\beta^2\cos\theta}{2\pi r^2\sin^3\theta}\right)^2
        - \frac{\beta\pa_\theta\beta \cos\theta}{2\pi r^4\sin^3\theta} + \frac{\beta^2\cos^2\theta}{\pi r^4\sin^4\theta} \ .
        \end{split}
        \label{coefficients_0}
        \enq

\subsubsection{Coefficients for $m \ne 0$}
For $m \ne 0$, the integrand given by equation (\ref{integrand_hyd_tor_m}) reduces to
        \beq
        {\cal E}_{\rm hyd} + {\cal E}_{\rm tor} = \frac{\beta^2}{8\pi r^2}\left[\pa_r(r R) + \pa_\theta S\right]^2
        + \frac{1}{2}\varpi^2(a_{m \ne 0} R^2 + b_{m \ne 0} R S + c_{m \ne 0} S^2) \ .
        \label{integrand_hyd_tor_min}
        \enq
The coefficients are given through (Tayler 1973; Goossens \& Veugelen 1978; Akg\"{u}n \& Wasserman 2008)
        \beq
        \begin{split}
        a_{m \ne 0} & = - \pa_r\varrho\pa_r\Phi - \frac{\varrho^2(\pa_r\Phi)^2}{\Gamma P}
        - \frac{\beta\pa_r\beta}{2\pi r^3\sin^2\theta} + \frac{m^2\beta^2}{4\pi r^4\sin^4\theta} \ , \\
        b_{m \ne 0} & = - \frac{\pa_r\varrho\pa_\theta\Phi}{r} - \frac{\pa_\theta\varrho\pa_r\Phi}{r}
        - \frac{2\varrho^2\pa_r\Phi\pa_\theta\Phi}{\Gamma P r}
        - \frac{\beta\pa_r\beta \cos\theta}{2\pi r^3\sin^3\theta}
        - \frac{\beta\pa_\theta\beta}{2\pi r^4\sin^2\theta} \ , \\
        c_{m \ne 0} & = - \frac{\pa_\theta\varrho\pa_\theta\Phi}{r^2} - \frac{\varrho^2(\pa_\theta\Phi)^2}{\Gamma P r^2}
        - \frac{\beta\pa_\theta\beta \cos\theta}{2\pi r^4\sin^3\theta} + \frac{m^2\beta^2}{4\pi r^4\sin^4\theta} \ .
        \end{split}
        \label{coefficients_ne0}
        \enq

\subsubsection{Proof that all continuous toroidal fields are unstable}\label{section_proof}
Tayler (1973) shows that a toroidal field with a non-zero current density on the axis is necessarily unstable, and that the instability occurs near the axis, regardless of field strength. Goossens, Biront \& Tayler (1981) further show that a toroidal field is unstable if there is some point in the star where the field strength is zero, but its derivative with respect to $\sin\theta$ is positive. Next, we show more generally that, in fact, all physically relevant toroidal fields are unstable (including the one presented in \S\ref{section_toroidal}, which is not covered by the previous arguments).

A toroidal field is unstable if, for some value of $m$, $c_m < 0$ somewhere in the star. For $m = 0$ (equation \ref{coefficients_0}), neglecting perturbations of the order of $B^4$ caused by the magnetic field, we have
        \beq
        c_0 = - \frac{\beta\pa_\theta\beta \cos\theta}{2\pi r^4\sin^3\theta}
        + \frac{\beta^2\cos^2\theta}{\pi r^4\sin^4\theta}
        = - \frac{\sin\theta\cos\theta}{4\pi r^4} \pa_\theta\left(\frac{\beta^2}{\sin^4\theta}\right) \ .
        \label{andreas_c0}
        \enq
On the other hand, for $m \ne 0$, we have, from equation (\ref{coefficients_ne0}),
        \beq
        c_{m \ne 0} = - \frac{\beta\pa_\theta\beta\cos\theta}{2\pi r^4 \sin^3\theta} + \frac{m^2\beta^2}{4\pi r^4 \sin^4\theta}
        = - \frac{\tan^{m^2-1}\theta\pa_\theta(\beta^2\cot^{m^2}\theta)}{4\pi r^4\sin^2\theta} \ .
        \label{andreas_cm}
        \enq
Considering specifically $m=1$, this simplifies to
        \beq
        c_1 = - \frac{\pa_\theta(\beta^2\cot\theta)}{4\pi r^4\sin^2\theta} \ .
        \label{andreas_c1}
        \enq

Now, we need to look at the behavior of $\beta$. In this case, the magnetic field and current density are $\m{B} = \beta\m\nabla\phi$ and $4\pi\m{J}/c = \m\nabla\beta\times\m\nabla\phi$, respectively (from equations \ref{pol_tor} and \ref{pol_tor_current}). At the axis, the magnetic field and current density cannot have $\hvarpi$ and $\hphi$ components, which implies that $\beta$ must go to zero faster than $\varpi \propto \sin\theta$. Then, as $\varpi \to 0$, we have $4\pi\m{J}/c \to \varpi^{-1}\pa_\varpi\beta \hz$. If we want the latter to be finite, we need $\beta$ to go to zero at least as fast as $\varpi^2 \propto \sin^2\theta$ as we approach the axis. Thus, in the above equations (for $m = 0$ and $m \ne 0$), $\beta^2$ easily cancels the singularities due to $\sin\theta$ at $\theta = 0$ and $\pi$.

In particular, consider the coefficient $c_0$ as given by equation (\ref{andreas_c0}). If $\beta \propto \sin^2\theta$ then $c_0$ vanishes everywhere, which, in the best case, implies marginal stability (if $b_0 = 0$ as well). If, on the other hand, $\beta$ goes to zero faster than $\sin^2\theta$ (as is the case for the field considered in this paper, for which $\beta = 0$ identically in a finite range of $\theta$), then the function $\beta^2/\sin^4\theta$ increases from zero to a finite value somewhere in the interval $0 < \theta < \pi/2$ (i.e.\ it has a positive derivative while $\cos\theta > 0$), and decreases from some finite value to zero for $\pi/2 < \theta < \pi$ (i.e.\ it has a negative derivative while $\cos\theta < 0$). Therefore, in both cases we will have some regions where $c_0 < 0$, thus leading to instability.

On the other hand, as can be seen from equation (\ref{andreas_c1}), we will have $c_1 < 0$ whenever the derivative $\pa_\theta(\beta^2\cot\theta)$ is positive. Note that $\beta^2\cot\theta = 0$ at $\theta = 0$, $\pi/2$, and $\pi$; $\beta^2\cot\theta \geqslant 0$ for $0 < \theta < \pi/2$, and $\beta^2\cot\theta \leqslant 0$ for $\pi/2 < \theta < \pi$. Thus, if $\beta$ is a continuous function, we will have $c_1 < 0$ somewhere in whichever hemisphere $\beta$ has some non-zero values. In other words, the $m = 1$ instabilities will happen in those regions where, moving on a spherical shell of constant radius in the direction of increasing $\theta$, $\beta^2\cot\theta$ increases from zero to some finite value while $\cot\theta > 0$, or from some finite negative value back to zero while $\cot\theta < 0$.

\subsubsection{Application to our particular magnetic field structure}
\begin{figure}
        \centerline{\includegraphics[scale=1]{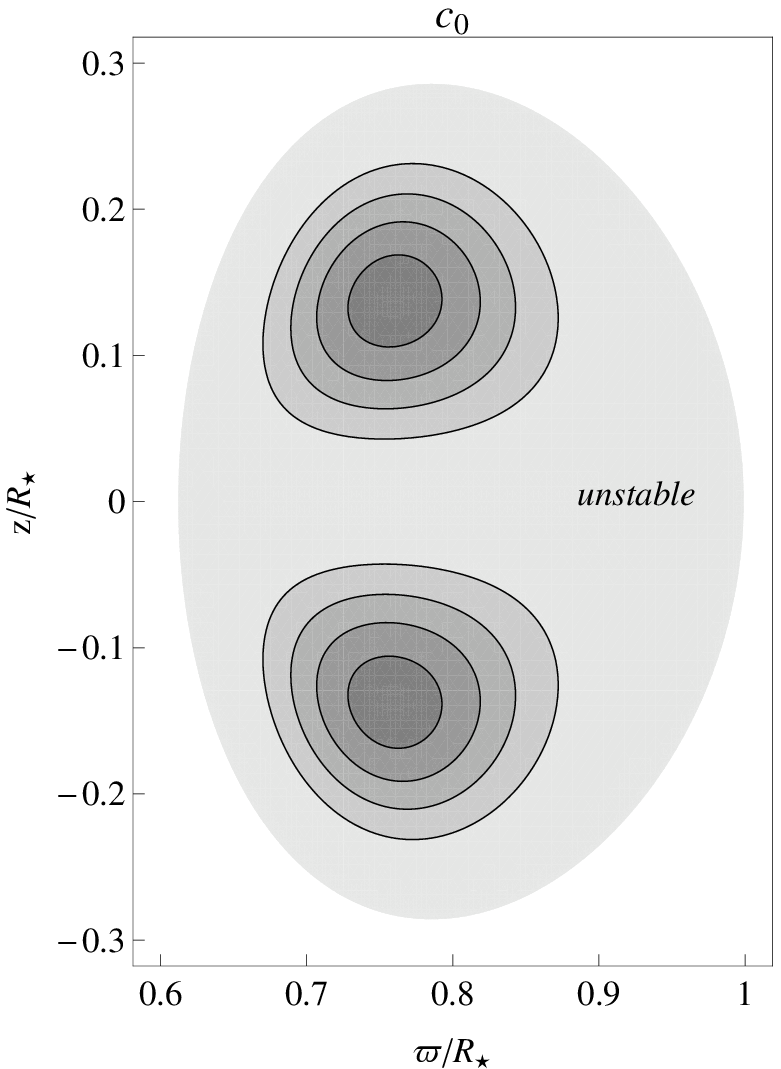} \ \includegraphics[scale=1]{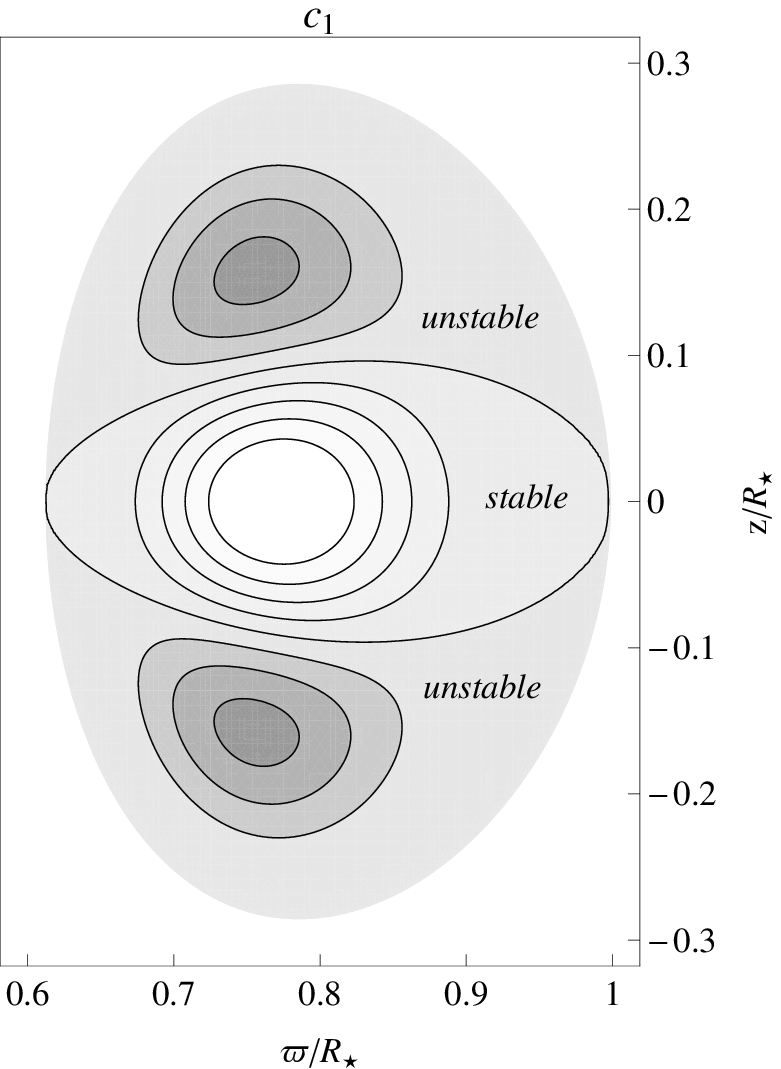}}
        \caption{Contours of Tayler's coefficients $c_0$ (equation \ref{coefficients_0}) and $c_1$ (equation \ref{coefficients_ne0}). The coefficients are shown in units of $b_{\rm tor}^2 B_{\rm o}^2/4\pi R_\star^2$. The equator (horizontal) and axis (vertical) are shown in units of stellar radii ($\varpi/R_\star$ and $z/R_\star$, respectively). The outer boundary is defined by the last poloidal field line that closes within the star (and is tangential to the surface at the equator); both $\beta = 0$ and $c_m = 0$ on this boundary. This defines the region where the toroidal field exists. The toroidal field is of the form given by equation (\ref{beta}) with $n = 2$. The contours are shown in the range $-1.2$ (dark) to $-0.3$ (light) for $c_0$, and $-0.9$ (darkest) to $1.2$ (white) for $c_1$ in increments of $0.3$. $c_m < 0$ implies instability. $c_0$ is zero along the equator ($\theta = \pi/2$) and on the boundary, and is negative everywhere else. The maximum of $c_1$ occurs along the equator, at $x \approx 0.772$; its minima are at $x \approx 0.772$ and $\theta \approx \pi/2 \pm 0.205$; its radial extent along the equator is $0.612 \lesssim x \leqslant 1$; and its angular extent in the meridional plane is $1.21 \lesssim \theta \lesssim 1.93$.}
        \label{fig_cm}
        \end{figure}

Now consider the application to our choice of toroidal magnetic field, where $\beta$ is given by equation (\ref{beta}), and $\alpha$ is given by equations (\ref{alpha}) and (\ref{dipole_func}). Keep in mind the renormalization of the functions $\alpha$ and $\beta$ carried out in accordance with equation (\ref{magnetic_field}) (we are now considering the case $\eta_{\rm pol} = 0$ and $\eta_{\rm tor} = 1$, so we have only a toroidal field, despite taking $\alpha \ne 0$). For completeness, in the next few lines, we will keep track of the power $n$ defined in equation (\ref{beta}). Thus, $\alpha(x,\theta) = f(x) \sin^2\theta$ and $\beta(x,\theta) = [\alpha(x,\theta) - 1]^n$ in the region where the toroidal field is non-zero. Note that this $\beta$ vanishes long before reaching the axis. We consider only the region where $\beta \ne 0$, i.e.\ $\alpha > 1$, or equivalently, $1/f(x) < \sin^2\theta \leqslant 1$. From equations (\ref{andreas_c0}) and (\ref{andreas_c1}), we have
        \beq
        \frac{c_{0,1}}{B_{\rm o}^2/4\pi R_\star^2} = - \frac{[f(x) \sin^2\theta - 1]^{2n - 1}}{x^4\sin^4\theta}G_{0,1} \ ,
        \label{g01}
        \enq
where $G_0(x,\theta) = 4\cos^2\theta[(n - 1) f(x)\sin^2\theta + 1]$ and $G_1(x,\theta) = f(x) \sin^2\theta (4n\cos^2\theta - 1) + 1$. In equation (\ref{g01}), the coefficient of $G_{0,1}$ is always non-positive, and the instability condition $c_{0,1} < 0$ thus requires $G_{0,1} > 0$. Note that for our particular choice of the magnetic field $G_0 \geqslant 0$, thus $c_0 \leqslant 0$, i.e.\ there is always an $m = 0$ instability. For $m = 1$, the instability condition $G_1 > 0$ is satisfied for $\sin^2\theta \approx 1/f(x)$, but is not satisfied for $\sin^2\theta \approx 1$, i.e.\ $c_1$ changes sign somewhere between these two values. The point where the sign change takes place is given as the real root of $G_1 = 0$,
        \beq
        \sin^2\theta_{\rm c} = \frac{1}{2}\left[1 - \frac{1}{4n}
        + \sqrt{\left(1 - \frac{1}{4n}\right)^2 + \frac{1}{nf(x)}}\right] \ .
        \enq
The unstable region is $1/f(x) < \sin^2\theta < \sin^2\theta_{\rm c}$, and the stable region is $\sin^2\theta_{\rm c} < \sin^2\theta \leqslant 1$. For $n = 2$ and $f(x) = f_{\rm max} \approx 1.14$ (\S\ref{section_poloidal}), corresponding to the largest extent in colatitude, the stable region is $83.2^\circ < \theta < 96.8^\circ$, and the rest of the interval where the toroidal field is present, $69.4^\circ < \theta < 110.6^\circ$ (\S\ref{section_toroidal}), is unstable. The contours of the coefficients $c_0$ and $c_1$ are shown in Fig. \ref{fig_cm}.

\subsubsection{Proof that the limiting case of perfect stable stratification implies stability}\label{section_perfect}
Typically, the hydrostatic force is much stronger than the magnetic force. This implies that, in a stably stratified star, any radial displacement will be acting against a prohibitively large buoyancy force. In the limiting case of \emph{perfect stable stratification}, let's consider a displacement field that is perpendicular to the restoring hydrostatic force, which to lowest order points in the radial direction (equation \ref{delta_f_hyd}). In addition, we require the fluid to be incompressible. In other words, the density remains constant as a fluid element is displaced (i.e.\ $\Delta\varrho = - \varrho\m\nabla\cdot\m\xi = 0$, while $\Delta P \ne 0$, implying that $\Gamma \to \infty$ from equation \ref{gamma_ad}). These two conditions, namely \emph{incompressibility} ($\m\nabla\cdot\m\xi = 0$) and \emph{orthogonality to the radial direction} ($\hr\cdot\m\xi = 0$), imply that the displacement field is described by a single unknown function (instead of three, as is the case for an unrestricted vector field). From these assumptions it also follows that the hydrostatic part of the energy vanishes (to first order in $B^2$, as in equation \ref{hydrostatic2}). Therefore, all we are left with is the variation in the magnetic energy. In our notation, the requirements $\m\nabla\cdot\m\xi = 0$ and $\hr\cdot\m\xi = 0$ correspond to setting $D_m = 0$ and $R = 0$ in the integrand of equation (\ref{integrand_hyd_tor}). To first order in $B^2$, the integrand reduces to (for any $m$)
        \beq
        {\cal E}_{\rm hyd} + {\cal E}_{\rm tor} = \frac{1}{8\pi r^2}\left[\beta^2(\pa_\theta S)^2
        + \frac{(m^2\beta^2 - \beta\pa_\theta\beta\sin 2\theta)S^2}{\sin^2\theta}\right] \ .
        \label{integrand_tor_perfect}
        \enq
This displacement field is restricted to such an extent that it is not possible to make the first positive definite term vanish by a suitable choice, unlike in the general case. Therefore, the coefficient of $S^2$ is no longer sufficient to assess stability. In fact, it is possible to show that the integral is always positive, which is not immediately obvious from the above form. It is made clearer by rewriting the integrand as
        \beq
        {\cal E}_{\rm hyd} + {\cal E}_{\rm tor} = \frac{1}{8\pi r^2}\left[\frac{(m^2-1)S^2\beta^2}{\sin^2\theta}
        + \beta^2(\pa_\theta S + S\cot\theta)^2 - \frac{\pa_\theta(S^2\beta^2\cos\theta)}{\sin\theta}\right] \ .
        \enq
The last term integrates to zero, since $S\sin\theta \to 0$ and $\beta/\sin\theta \to 0$ on the symmetry axis. On the other hand, the sum of the first two terms is always positive for $m^2 \geqslant 1$. For $m = 0$, the only displacement field consistent with $\m\nabla\cdot\m\xi = 0$ and $\hr\cdot\m\xi = 0$ that does not diverge is of the form $\m\xi = \xi_\phi(r,\theta)\hphi$, which has no effect on the toroidal field (equation \ref{Faraday}). Thus, we conclude that the equilibrium is always stable to this very restricted set of perturbations, as previously noted by Dicke (1979). Therefore, in order to obtain instabilities, the restrictions due to perfect stable stratification must be relaxed. Intuitively, this is reasonable, because without a radial displacement it is not possible to have either a global or a local interchange of toroidal field lines (corresponding to the previously identified ``interchange'' or ``kink'' instabilities).

\subsubsection{Constraints on the destabilizing displacement field}\label{section_destabilize}
In this section we will consider the properties of the displacement field that destabilizes the toroidal magnetic field. Some simple observations can be inferred by noting that the energy integrand can be written as the sum of a positive definite term and a quadratic as in equation (\ref{integrand_hyd_tor_m}). We would like the quadratic to be negative and the positive definite term to be as small as possible.

First, consider the quadratic, $Q(R,S) \equiv a_m R^2 + b_m R S + c_m S^2$. Since $a_m$ is always large and positive, and $c_m$ is small and negative in some region, $|R|$ must be small compared to $|S|$ in order to allow the energy to become negative. On the other hand, $R$ cannot be zero, as that reduces the integrand in equation (\ref{integrand_hyd_tor_min}) to the form given by equation (\ref{integrand_tor_perfect}), which was shown to be always positive. This is because the integrand in equation (\ref{integrand_hyd_tor_min}) is already minimized with respect to $T$, i.e.\ we have implicitly substituted the condition given by equation (\ref{minimum_T}), which implies that if $R = 0$ then $D_m \approx 0$, thus leading us back to equation (\ref{integrand_tor_perfect}) for the perfect stable stratification. Thus, \emph{$|R|$ must be small compared to $|S|$, but non-zero}. Since $a_m > 0$, $Q(R,S)$ can be minimized with respect to $R$. The minimum is given by
        \beq
        R_{\rm min} = - \frac{b_m}{2a_m} S \mtext{and}
        Q_{\rm min} = \left(-\frac{b_m^2}{4a_m} + c_m\right)S^2 \ .
        \label{minimize_Q}
        \enq
For instability, we must have $Q_{\rm min} < 0$, implying that $b_m^2 > 4 a_m c_m$. Since typically $|b_m| \sim |c_m| \ll a_m$ (equation \ref{leading_am_cm}), this will be satisfied when $c_m < 0$ (plus a thin region where $c_m \geqslant 0$ but very small), and we will have $Q_{\rm min} \approx c_m S^2$. If $c_m < 0$ is confined to a region of size $(\Delta r,\Delta\theta)$ in the relevant coordinates, the displacement field should also be roughly confined to this region, as there would otherwise be a positive contribution to the energy from the region where $c_m > 0$. Thus, in particular, $|\pa_\theta S| \gtrsim |S| / \Delta\theta$ (since $S$ must vanish near the boundary of the region, we have $|\Delta S| \sim |S|$).

The positive definite terms for $m = 0$ and $m \ne 0$ are significantly different, and the two cases need to be treated separately. First, we will consider the case $m \ne 0$. Since the region we are interested in is near the equator (i.e.\ $\sin\theta \approx 1$ and $\cos\theta \sim \Delta\theta$), we have $c_{m \ne 0} \sim - \beta^2 / 4\pi r^4$ (equation \ref{andreas_c1}). For instability, we need ${\cal E}_{\rm hyd} + {\cal E}_{\rm tor} < 0$ in equation (\ref{integrand_hyd_tor_min}), which implies $\left[\pa_r(r R) + \pa_\theta S\right]^2 \lesssim S^2$. Thus, we need $|\pa_r(r R) + \pa_\theta S| \lesssim |S| \lesssim |\pa_\theta S| \Delta\theta$, using the above inequality. In other words, the more confined the displacement field is in latitude (i.e.\ the smaller the range $\Delta\theta$, forced by the condition $c_{m \ne 0} < 0$), the more precisely the two derivatives on the left-hand side need to cancel each other. In the limit $\Delta\theta \to 0$, it is necessary to enforce
        \beq
        \pa_r(r R) + \pa_\theta S = 0 \ .
        \label{equation_mnot0}
        \enq

In addition, from the confinement to an interval $\Delta r$ we have $|\pa_r(r R)| \gtrsim r |R| / \Delta r$, and from the cancelation with $\pa_\theta S$ we have $|\pa_r(r R)| \gtrsim |S| / \Delta\theta$. The ratio of these two lower bounds is, using equations (\ref{leading_am_cm}) and (\ref{minimize_Q}), $\mathfrak{R} = (|R| / |S| ) (r\Delta\theta / \Delta r) \approx (|b_{m \ne 0}| / 2 a_{m \ne 0}) (r\Delta\theta / \Delta r) \sim (B^2 / 8\pi P_0) (r\Delta\theta / \Delta r) \ll 1$. Thus, the lower bound on $|\pa_r(r R)|$ from the requirement of canceling $\pa_\theta S$ (even if not exactly) is much larger than the bound from being confined to the region $(\Delta r,\Delta\theta)$. Thus, the length scale of variation of $R$ must be $\delta_r \ll \Delta r$. Note that this does not mean that $\m{\xi}$ is confined to a region as thin as $\delta_r$. It could extend over the whole $\Delta r$, but it would have to oscillate on a radial length scale $\delta_r$.

For $m = 0$ the coefficient of the positive definite term in equation (\ref{integrand_hyd_tor_m}) is much larger than in the $m \ne 0$ case, $K_0 \gg K_{m \ne 0}$. This implies that the quantity in parentheses must cancel out even more precisely. Keeping only leading order terms, we get
        \beq
        \frac{\pa_r(r^3 R)}{r^3} + \frac{\pa_\theta(S \sin^2\theta)}{r \sin^2\theta}
        + \frac{\pa_r P_0}{\Gamma P_0} R = 0 \ .
        \enq
This is different from equation (\ref{equation_mnot0}) for the $m \ne 0$ case. In what follows we will consider only the simpler case of $m \ne 0$, and the $m = 0$ case will be left for future work.

\subsubsection{Particular displacement field for $m \ne 0$}\label{section_particular}
As discussed in the previous section, the quadratic part of the integrand for $m \ne 0$ (equation \ref{integrand_hyd_tor_min}) can be made negative by a suitable choice of the amplitudes of the functions $R$ and $S$. In addition, the first term, which is positive definite, can be minimized by a suitable choice of the derivatives of these functions. In particular, the best choice might be when this term is made to vanish (Goossens \& Tayler 1980; Goossens \& Biront 1980), which leads to the condition given by equation (\ref{equation_mnot0}). This equation is satisfied by solutions of the form
        \beq
        R = \frac{\pa_\theta\Pi}{x} \mtext{and} S = - \pa_x\Pi \ ,
        \label{displacement_RS}
        \enq
where $x$ is the dimensionless radial coordinate $x = r/R_\star$, and $\Pi(x,\theta)$ is some scalar generating function for the displacement field. While we can choose $R$ and $S$ so that the positive definite term vanishes, $T$ has a particular value for which the integrand in equation (\ref{integrand_hyd_tor}) is minimized. This value of $T$ is expressible in terms of $R$, $S$, and their derivatives, corresponding to the condition given by equation (\ref{minimum_T}). Using equations (\ref{Lambda_D}) and (\ref{equation_mnot0}), and keeping only the lowest order terms, we have
        \beq
        \frac{m T}{r\sin\theta} = D_0 - \frac{\varrho\Lambda(\Phi)}{\Gamma P}
        \approx \left(\frac{2}{r} + \frac{\pa_r P_0}{\Gamma P_0}\right)R
        + \left(\frac{2\cot\theta}{r}\right)S \ .
        \label{displacement_T}
        \enq

As noted in \S\ref{section_stratification}, to lowest order, we can drop all magnetic corrections to the background quantities in the hydrostatic part of the energy (equation \ref{calE_hyd}), which then becomes, using the equation of equilibrium (equation \ref{euler}), the definition of $\gamma$ (equation \ref{gamma}), and the value of $D_m$ from equation (\ref{minimum_T}),
        \beq
        {\cal E}_{\rm hyd} \approx \frac{1}{2}\left(\frac{1}{\gamma} - \frac{1}{\Gamma}\right)
        \frac{(\pa_r P_0)^2}{P_0}R^2 r^2\sin^2\theta \approx \frac{1}{2} a_{m \ne 0} R^2 r^2 \sin^2\theta \ .
        \label{leading_hyd}
        \enq
The last equality follows from equation (\ref{leading_am_cm}). In a stably stratified star $\Gamma > \gamma$, so that the hydrostatic part is always positive. The toroidal part of the energy follows from equation (\ref{calE_mag}), to leading order and using $|R| \ll |S|$,
        \beq
        {\cal E}_{\rm tor} \approx \frac{1}{2} c_{m \ne 0} S^2 r^2 \sin^2\theta \ .
        \label{leading_tor}
        \enq
Thus, the total integrand given by equation (\ref{integrand_hyd_tor_min}) reduces to ${\cal E}_{\rm hyd} + {\cal E}_{\rm tor} \approx \frac{1}{2}(a_{m \ne 0} R^2 + c_{m \ne 0} S^2) r^2 \sin^2\theta$. Since $|R| \ll |S|$ and $a_{m \ne 0} \gg |b_{m \ne 0}| \sim |c_{m \ne 0}|$ (equation \ref{leading_am_cm}), it follows that the $b_{m \ne 0}RS$ term in the quadratic can be dropped.

        \begin{figure}
        \centerline{\includegraphics[scale=0.8]{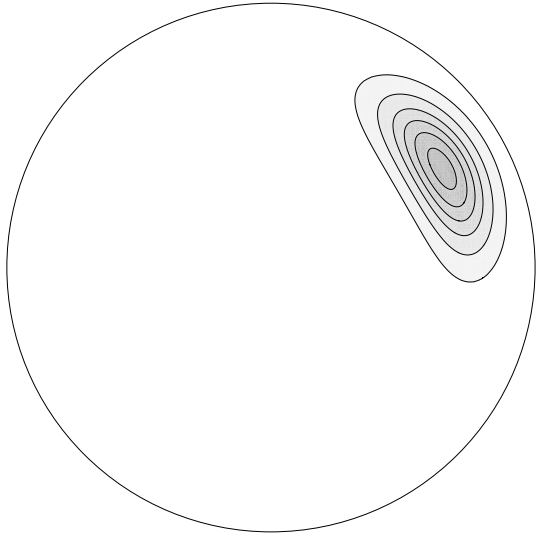}}
        \caption{Contours of a greatly exaggerated sample generating function $\Pi$ for the displacement field (equation \ref{pi}). The contours are the streamlines of the displacement field. The stellar surface is shown as a solid line. The values of the various parameters used in this plot are $x_0 = 3/4$, $\theta_0 = \pi/3$, $\delta_r = 1/5$, $\delta_\theta = \pi/5$, and $\sigma = 3$. The actual generating function used in our calculations is much more confined.}
        \label{fig_displacement}
        \end{figure}

We can now proceed to construct a particular displacement field that will make the purely toroidal magnetic field unstable. We will assume that the displacement field is confined to a region within the star and is zero everywhere else. In order to prove that the stability conditions are both sufficient and necessary, Tayler (1973) assumes a particular solution of the form $\Pi(x,\theta) \propto \sin kx\sin\ell\theta$ in a finite volume, bounded by a surface on which $\Pi(x,\theta) = 0$. This corresponds to a finite displacement field which is tangential to the boundaries. While this form is acceptable for a purely toroidal field (since both $\m\xi$ and $\m{B}$ are tangential to the surface), in our case we will eventually incorporate a poloidal field as well, and any discontinuity in the displacement field at the boundaries would cause divergences (cutting the field lines). In order to avoid such pathologies, we would therefore like $\m\xi$ to go to zero at the boundary, and its derivatives to remain finite everywhere. Thus, we choose
        \beq
        \Pi(x,\theta) = \frac{\xi_{\rm o}}{R_\star}\left[1 - \chi^2(x,\theta)\right]^\sigma \ , \mtext{where}
        \chi^2(x,\theta) = \frac{(x - x_0)^2}{\delta_r^2}
        + \frac{(\theta - \theta_0)^2}{\delta_\theta^2} \ .
        \label{pi}
        \enq
The factor $\xi_{\rm o}/R_\star$ sets the amplitude of the displacement field (equation \ref{xi}). The displacement field is zero on the boundary (defined by $\chi = 1$) and outside of it. This particular choice of $\chi$ corresponds to a donut-shaped region with a meridional cross-section in the shape of a distorted ellipse (Fig. \ref{fig_displacement}). In order for the derivatives of the displacement field to remain finite, we must have $\sigma \geqslant 2$.

As shown in \S\ref{section_proof}, the second condition in equation (\ref{conditions}) is always violated somewhere for any non-singular toroidal magnetic field. Therefore, we will have $c_{m \ne 0} < 0$ in some region (Fig. \ref{fig_cm}). If we choose the displacement field to be confined near the minimum of $c_{m \ne 0}$ and make $c_{m \ne 0} S^2$ to be the dominant term in the quadratic in equation (\ref{integrand_hyd_tor_min}), then the energy will be negative. Thus, we want, roughly, $a_{m \ne 0} R^2 \lesssim |c_{m \ne 0}| S^2$. For a displacement field given by equations (\ref{displacement_RS}) and (\ref{pi}), this implies that we must have $|R|/|S| \sim \delta_r / \delta_\theta \lesssim \sqrt{|c_{m \ne 0}|/a_{m \ne 0}} \sim B/\sqrt{P_0} \ll 1$. Here, we have used $|x - x_0| \leqslant \delta_r$, $|\theta - \theta_0| \leqslant \delta_\theta$, and equation (\ref{leading_am_cm}). Since $\delta_\theta$ cannot be much larger than the angular extent of the negative region of the coefficient $c_{m \ne 0}$ (Fig. \ref{fig_cm}), this then imposes a very stringent upper limit on $\delta_r$.

\subsection{Stability of a toroidal field in the presence of a weaker poloidal component}
The displacement field constructed in the previous section makes the sum of the hydrostatic and toroidal parts of the energy negative, thus leading to an instability. In the present section we will consider the case when a weaker poloidal component is added. This poloidal field will give an additional positive contribution to the energy and will help stabilize the instability of the toroidal field. Our goal is to determine the minimum strength of the poloidal field relative to the toroidal field in order to achieve stability. Note that this treatment inherently relies on the implicit assumption that the poloidal field is sufficiently weaker than the toroidal field, so that the displacement field discussed in the previous section is still close to being the most unstable mode. This is not obviously true, but appears to be validated by the eventual results.

\subsubsection{Leading order estimates of the energy terms}
Here, we will give estimates of the hydrostatic, toroidal, and poloidal parts of the energy in terms of the parameters of the particular displacement field constructed in the previous section. Since our $R$, $S$, and $T$ are real functions, the cross term in equation (\ref{calE_mag}) has no real part and is physically irrelevant. The total energy in terms of the integrand ${\cal E}$ is (from equation \ref{energy_pert}), carrying out the integration over $\phi$,
        \beq
        \delta W = \delta W_{\rm hyd} + \delta W_{\rm tor} + \delta W_{\rm pol}
        = \frac{1}{2}\int{\cal E}dV = \pi \int {\cal E} r^2\sin\theta dr d\theta \ .
        \label{exact_int}
        \enq

The hydrostatic and toroidal integrands are given to leading order in $B^2/8\pi P_0$ by equations (\ref{leading_hyd}) and (\ref{leading_tor}). On the other hand, the poloidal integrand is given by equation (\ref{calE_mag}). Here, we need to make use of the azimuthal displacement, which is related to the other two components by equation (\ref{displacement_T}). Since $|R| \ll |S|$ (or $\delta_r \ll \delta_\theta < 1$), to leading order we have $m T \approx 2S\cos\theta$. Thus, $R \propto 1/\delta_\theta$, $S \propto 1/\delta_r$, and $T \propto 1/\delta_r$. Each subsequent derivative $\pa_x$ of the displacement field brings in an additional factor of $1/\delta_r$, and similarly, $\pa_\theta$ brings in a factor of $1/\delta_\theta$. It then follows that the four terms in the poloidal integrand as given by equation (\ref{calE_mag}) scale as $1/\delta_r^4$, $1/\delta_r^2\delta_\theta^2$, $1/\delta_r^4$, and $1/\delta_r^2$, respectively. We thus conclude that the first and third terms in the poloidal integrand are the largest, followed by the second term, while the fourth term, which is also the only negative term in the expression, is the smallest. Thus, it becomes obvious that, for the displacement field of the form constructed here, the poloidal contribution is \emph{overwhelmingly} positive. To leading order, keeping only the first and third terms of the poloidal integrand (equation \ref{calE_mag}), we have
        \beq
        {\cal E}_{\rm pol} \approx \frac{(\pa_\theta\alpha\pa_r S)^2}{8\pi r^2}\left(1 + \frac{4\cos^2\theta}{m^2}\right) \ .
        \enq
We define a new coefficient for the poloidal field, in analogy to the coefficients $a_{m \ne 0}$ and $c_{m \ne 0}$,
        \beq
        d_{m \ne 0} \equiv \frac{9}{4\pi}\left(\frac{\pa_\theta\alpha}{r^3\sin\theta}\right)^2\left(1 + \frac{4\cos^2\theta}{m^2}\right) \ .
        \enq
Thus, the hydrostatic, toroidal, and poloidal energies can be written as (using equations \ref{leading_hyd} and \ref{leading_tor}, and the above definitions)
        \beq
        \begin{split}
        \delta W_{\rm hyd} & \approx \frac{\pi}{2} \int a_{m \ne 0} R^2 r^4\sin^3\theta dr d\theta \ , \\
        \delta W_{\rm tor} & \approx \frac{\pi}{2} \int c_{m \ne 0} S^2 r^4\sin^3\theta dr d\theta \ , \\
        \delta W_{\rm pol} &\approx \frac{\pi}{2} \int d_{m \ne 0} \left(\frac{r \pa_r S}{3}\right)^2 r^4\sin^3\theta dr d\theta \ .
        \end{split}
        \label{approx_int}
        \enq
These approximations are remarkably accurate for the typical values of the parameters discussed in \S\ref{section_particular} (the errors with respect to the exact integrals are of the order of $10^{-6}$ or less).

In the limiting case of a vanishing area of integration (i.e.\ as $\delta_r \to 0$ and $\delta_\theta \to 0$ simultaneously), all slowly varying quantities can be taken as constant, and can be evaluated at the center of the displacement field $(x_0,\theta_0)$. In other words, the integrations can be carried out over the rapidly varying functions of the variables $(x-x_0)/\delta_r$ and $(\theta-\theta_0)/\delta_\theta$, and all other slowly varying functions of $x$ and $\theta$ can be taken out of the integrations. We thus have
        \beq
        \begin{split}
        \delta W_{\rm hyd} & \to \frac{\pi R_\star^5}{2} a_{m \ne 0}(x_0,\theta_0) x_0^2\sin^3\theta_0 \int (\pa_\theta\Pi)^2 dx d\theta \ , \\
        \delta W_{\rm tor} & \to \frac{\pi R_\star^5}{2} c_{m \ne 0}(x_0,\theta_0) x_0^4\sin^3\theta_0 \int (\pa_x\Pi)^2 dx d\theta \ , \\
        \delta W_{\rm pol} & \to \frac{\pi R_\star^5}{2} d_{m \ne 0}(x_0,\theta_0) x_0^6\sin^3\theta_0 \int \frac{(\pa_x^2\Pi)^2}{9} dx d\theta \ .
        \end{split}
        \label{limit_int}
        \enq
The remaining integrations can be carried out in polar coordinates through the substitutions $(x-x_0)/\delta_r = \chi\cos\psi$ and $(\theta-\theta_0)/\delta_r = \chi\sin\psi$, yielding
        \beq
        \begin{split}
        I_\sigma & \equiv \frac{R_\star^2}{\xi_{\rm o}^2} \frac{\delta_\theta}{\delta_r} \int (\pa_\theta\Pi)^2 dx d\theta
        = \frac{R_\star^2}{\xi_{\rm o}^2} \frac{\delta_r}{\delta_\theta} \int (\pa_x\Pi)^2 dx d\theta = \frac{\pi\sigma}{2(\sigma - 1/2)} \ , \\
        J_\sigma & \equiv \frac{R_\star^2}{\xi_{\rm o}^2} \frac{\delta_r^3}{\delta_\theta} \int \frac{(\pa_x^2\Pi)^2}{9} dx d\theta
        = \frac{\pi\sigma^2(\sigma - 1)}{6(\sigma - 1/2)(\sigma - 3/2)} \ .
        \end{split}
        \label{IJ}
        \enq
As noted following equation (\ref{pi}), we are interested in the case $\sigma \geqslant 2$. With these definitions, the limiting forms of the energy perturbations can be written as
        \beq
        \begin{split}
        \delta W_{\rm hyd} & \to (\Gamma/\gamma - 1) P_c \xi_{\rm o}^2 R_\star k_{\rm hyd}^{} I_\sigma \frac{\delta_r}{\delta_\theta} \ , \\
        \delta W_{\rm tor} & \to - \frac{B_{\rm o}^2\xi_{\rm o}^2 R_\star}{8\pi} b_{\rm tor}^2 k_{\rm tor}^{} I_\sigma \frac{\delta_\theta}{\delta_r}\ , \\
        \delta W_{\rm pol} & \to \frac{B_{\rm o}^2\xi_{\rm o}^2 R_\star}{8\pi} b_{\rm pol}^2 k_{\rm pol}^{} J_\sigma \frac{\delta_\theta}{\delta_r^3} \ ,
        \end{split}
        \label{coeff_acd}
        \enq
where $k_{\rm hyd}$, $k_{\rm tor}$ and $k_{\rm pol}$ are numerical constants which are independent of $\delta_r$, $\delta_\theta$, and $\sigma$, and whose values are given in Table \ref{table_coeff}. The factor $\Gamma/\gamma - 1$ (also evaluated at the same point as the coefficients) has also been explicitly written in the hydrostatic part in order to keep track of the dependence on stable stratification. The amplitudes of the toroidal and poloidal fields are set by $b_{\rm tor}$ and $b_{\rm pol}$ as defined through equation (\ref{amplitudes}). In addition, we have explicitly factored out all dimensional quantities $\xi_{\rm o}$, $P_c$, $B_{\rm o}$, and $R_\star$. The convergence of the hydrostatic, toroidal and poloidal energies to the limiting values of the approximations is demonstrated in Fig. \ref{fig_coeff}.

        \begin{table}
        \caption{Numerical values of the dimensionless coefficients $k_{\rm hyd}$, $k_{\rm tor}$, and $k_{\rm pol}$, defined through equation (\ref{coeff_acd}), for $x_0 = 0.772$ and $\theta_0 = 1.37$, corresponding to the minimum of the coefficient $c_1$ (Fig. \ref{fig_cm}). The background quantities are taken as in \S\ref{section_magnetic}, with $p = 8\pi P_c/B_{\rm o}^2 = 10^6$, and the magnetic field structure is that of \S\ref{section_poloidal} and \S\ref{section_toroidal}. We have also set $m = 1$ and $\Gamma/\gamma - 1 = 1/4$.}
        \label{table_coeff}
        \center{\begin{tabular}{cc}
        \hline Coefficient & Value \\ \hline
        $k_{\rm hyd}$ & 2.70 \\
        $k_{\rm tor}$ & 1.06 \\
        $k_{\rm pol}$ & 0.0832 \\
        \hline
        \end{tabular}}
        \end{table}

\subsubsection{Stability criteria}\label{section_criteria}
For stability, we must have $\delta W = \delta W_{\rm hyd} + \delta W_{\rm tor} + \delta W_{\rm pol} > 0$. Using equation (\ref{coeff_acd}) this can be written as, noting that $k_{\rm tor}$ is defined as a positive number and dropping common factors,
        \beq
        (\Gamma/\gamma - 1) p k_{\rm hyd}^{} I_\sigma \frac{\delta_r}{\delta_\theta}
        - b_{\rm tor}^2 k_{\rm tor}^{} I_\sigma \frac{\delta_\theta}{\delta_r}
        + b_{\rm pol}^2 k_{\rm pol}^{} J_\sigma \frac{\delta_\theta}{\delta_r^3} > 0 \ .
        \enq
Here, we have defined $p \equiv 8\pi P_c/B_{\rm o}^2$. We can rewrite this inequality as a lower bound on the amplitude of the poloidal field relative to the toroidal field,
        \beq
        \left(\frac{b_{\rm pol}}{b_{\rm tor}}\right)^2 > \delta_\theta^2 \frac{I_\sigma}{J_\sigma} \left[\frac{k_{\rm tor}}{k_{\rm pol}}
        \left(\frac{\delta_r}{\delta_\theta}\right)^2
        - \frac{k_{\rm hyd}}{k_{\rm pol}} \frac{(\Gamma/\gamma - 1) p}{b_{\rm tor}^2} \left(\frac{\delta_r}{\delta_\theta}\right)^4 \right] \ .
        \label{frac_b}
        \enq
We are interested in finding the \emph{minimum} poloidal field strength that can stabilize the magnetic field against all possible perturbations. Therefore, we would like to \emph{maximize} the expression on the right-hand side with respect to the various parameters involved ($\sigma$, $\delta_r$ and $\delta_\theta$). First, the ratio $I_\sigma/J_\sigma$ can be maximized with respect to $\sigma$, which for $\sigma \geqslant 2$ gives $\sigma = (3 + \sqrt{3})/2 \approx 2.37$ and $I_\sigma/J_\sigma = 3(2-\sqrt{3}) \approx 0.804$ (see equation \ref{IJ}). Next, the expression in square brackets can be maximized with respect to the ratio $\delta_r/\delta_\theta$ (or, equivalently, with respect to $\delta_r$ while keeping $\delta_\theta$ constant), yielding
        \beq
        \left(\frac{\delta_r}{\delta_\theta}\right)^2 = \frac{k_{\rm tor}^{}}{2 k_{\rm hyd}^{}} \frac{b_{\rm tor}^2}{(\Gamma/\gamma - 1) p} \ .
        \label{frac_delta}
        \enq
Note that for this value we indeed have $\delta W_{\rm hyd} + \delta W_{\rm tor} < 0$, as is required for the instability of the purely toroidal field in the first place. Setting $b_{\rm tor} = 1$ (which corresponds to measuring the poloidal field strength in terms of the toroidal one), $\Gamma/\gamma - 1 = 1/4$, $p = 10^6$, and using the tabulated values of the coefficients from Table \ref{table_coeff}, we obtain $\delta_r/\delta_\theta \approx 10^{-3}$.

\begin{figure}
        \centerline{\includegraphics[scale=1]{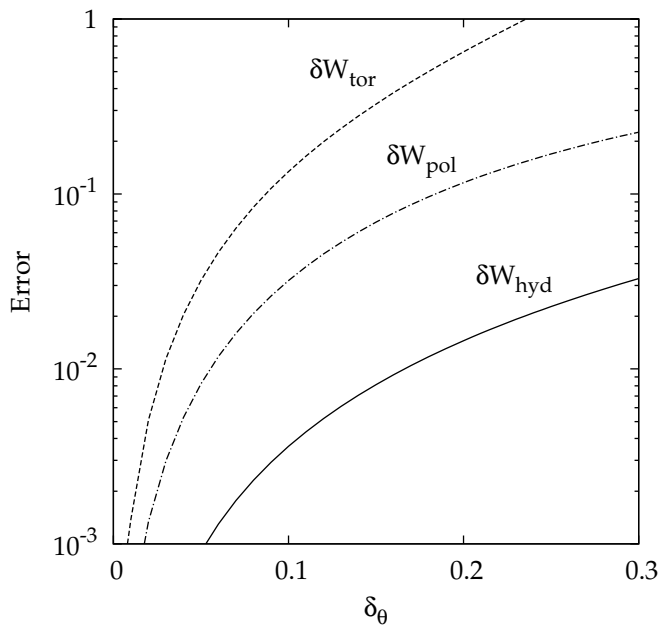}}
        \caption{Errors between the limiting values (equation \ref{coeff_acd}) and the exact values of the hydrostatic, toroidal and poloidal energies, as functions of $\delta_\theta$, where error = (limiting value - exact value)/exact value. All parameters, background quantities, and coefficients involved in the calculations are taken as in Table \ref{table_coeff}. In addition, $\delta_r$ and $\sigma$ are taken to have the values that minimize the total energy (or, equivalently, that maximize the ratio $b_{\rm pol}/b_{\rm tor}$, as discussed in the text following equation \ref{frac_b}), and $\delta_\theta$ is allowed to vary. The exact values converge to the limiting values as $\delta_\theta$ decreases.}
        \label{fig_coeff}
        \end{figure}

The remaining dependence of $(b_{\rm pol}/b_{\rm tor})^2$ on $\delta_\theta$ is monotonically increasing, therefore we need to evaluate the largest physically reasonable value of this parameter. For a displacement field centered at the minimum of the coefficient $c_1$, this value cannot be much larger than the angular extent of the negative region of the coefficient (Fig. \ref{fig_cm}). Otherwise, the toroidal energy $\delta W_{\rm tor}$ could no longer be made negative. In fact, for any sufficiently small $\delta_r$ ($\delta_r \lesssim 10^{-2}$) this condition (namely, $\delta W_{\rm tor} < 0$) translates into $\delta_\theta < 0.4$. In fact, we will take the largest value of this parameter to be about $\delta_\theta \approx 0.24$, which also corresponds to where our approximations start to fail (at this point the relative error between the exact and limiting values for the toroidal energy reaches $100\%$, corresponding to a factor of 2 error, Fig. \ref{fig_coeff}).

Plugging equation (\ref{frac_delta}) into equation (\ref{frac_b}), and using the values of the parameters and coefficients discussed here and in Table \ref{table_coeff}, while explicitly keeping track of the dependence on $b_{\rm tor}$, $\Gamma/\gamma - 1$ and $p$, we obtain
        \beq
        \left(\frac{b_{\rm pol}}{b_{\rm tor}}\right)^2 > \delta_\theta^2 \frac{I_\sigma}{J_\sigma}
        \frac{k_{\rm tor}^2}{4 k_{\rm hyd} k_{\rm pol}} \frac{b_{\rm tor}^2}{(\Gamma/\gamma - 1) p}
        \approx 5.8 \times 10^{-2} \frac{b_{\rm tor}^2}{(\Gamma/\gamma - 1) p} \ .
        \label{frac_b2}
        \enq
Since $p/b_{\rm tor}^2 \gg 1$, this is a very small lower bound on the amplitude of the poloidal field needed to stabilize the toroidal field, thus effectively justifying our treatment of the poloidal field as small in comparison to the toroidal field. We can rewrite this result in terms of the energies stored in the magnetic and gravitational fields. From equations (\ref{energy_bind}) and (\ref{energy_mag}), we have
        \beq
        \frac{E_{\rm pol}}{E_{\rm tor}} \approx 4.3 \left(\frac{b_{\rm pol}}{b_{\rm tor}}\right)^2 \mtext{and}
        \frac{E_{\rm tor}}{E_{\rm grav}} \approx 6.7 \times 10^{-2} \frac{b_{\rm tor}^2}{p} \ .
        \enq
Replacing these in equation (\ref{frac_b2}), we get
        \beq
        \frac{E_{\rm pol}}{E_{\rm tor}} \gtrsim \frac{3.7}{\Gamma/\gamma - 1} \frac{E_{\rm tor}}{E_{\rm grav}} \ .
        \label{result}
        \enq

Observations provide us with an upper limit on the poloidal magnetic field strength. The above equation then gives us an upper limit on the toroidal field strength. Taking $E_{\rm pol} / E_{\rm grav} < 10^{-6}$, we get $E_{\rm tor} / E_{\rm grav} < 5.2 \times 10^{-4} \sqrt{\Gamma/\gamma - 1}$. Thus, the maximum toroidal field strength depends on how stably stratified the star is through the factor $\Gamma/\gamma - 1$. The more stably stratified the star, the stronger the maximum toroidal field for a given poloidal field strength. For main-sequence stars, $\Gamma/\gamma - 1 \approx 1/4$ and we obtain $E_{\rm tor} / E_{\rm grav} < 2.6 \times 10^{-4}$, while for neutron stars, $\Gamma/\gamma - 1 \sim 10^{-2}$ and we have $E_{\rm tor} / E_{\rm grav} \lesssim 5 \times 10^{-5}$. (The estimates of $\Gamma/\gamma-1$ are discussed in Reisenegger 2009.) This also implies that a significant portion of the magnetic energy may be hidden in the toroidal field, while only the poloidal field is observed.

In particular, we can apply this result to the case of magnetars. For a $1.4 M_\odot$ neutron star with 10\,km radius, the gravitational energy is $E_{\rm grav} \approx 4\times 10^{53}$\,erg (equation \ref{energy_bind}). The energy of the poloidal field is $E_{\rm pol} \approx 2\times 10^{48} B_{15}^2$\,erg (equation \ref{energy_mag}), where $B_{15}$ is the surface magnetic field strength \emph{at the equator} in units of $10^{15}$\,G. (Note that, given the existence of closed poloidal field lines, this number is an order of magnitude higher than the most naive estimate obtained by multiplying the energy density corresponding to the surface field, $B^2/8\pi$, by the volume of the star.) Using $\Gamma/\gamma - 1 \sim 10^{-2}$, we then obtain an upper limit for the energy of the toroidal field as $E_{\rm tor} \lesssim 5\times 10^{49} B_{15}$\,erg. Note the linear dependence of the maximum toroidal energy on the surface magnetic field strength. The maximum toroidal field strength is then given through (using equations \ref{amplitudes} and \ref{energy_mag})
        \beq
        \left(B_{\rm tor}\right)_{\rm max} \lesssim 10^{17} B_{15}^{1/2} \mbox{G} \ .
        \label{max_tor}
        \enq

Soft gamma repeaters (SGRs) release as much as a few $10^{46}$\,erg energy in a single outburst (Mereghetti 2008). If the outbursts are repeated every century or so over a period of 10 millennia, then the total energy release is of the order of a few $10^{48}$\,erg. These could only be explained in terms of magnetars with poloidal fields in the excess of $10^{15}$\,G. However, inclusion of a toroidal field increases the potential energy available for outbursts, and a lower surface magnetic field of the order of $10^{14}$\,G would be sufficient to explain the observations.

A particularly interesting case is that of SGR 0418+5729, where the inferred surface magnetic field strength is just below $10^{13}$\,G (Rea et al.\ 2010). Its observed X-ray luminosity is $\sim 6.2\times 10^{31}$\,erg/s, and its characteristic (spin-down) age is $\sim 2.4\times 10^7$\,yr. If the object is assumed to maintain the same level of activity throughout its life, then the total energy required would be $\sim 5\times 10^{46}$\,erg. Using the formulae of the preceding paragraph, the energy content of a $10^{13}$\,G poloidal field is $E_{\rm pol} \approx 2\times 10^{44}$\,erg, i.e.\ more than two orders of magnitude less than what is required. On the other hand, the maximum allowed toroidal field energy in this case is $E_{\rm tor} \lesssim 5\times 10^{47}$\,erg, which would be quite sufficient.

Of course, using such a toroidal field reservoir to explain magnetar energetics requires the magnetic energy to be released on a timescale comparable with the magnetar lifetime. The mechanism for this might be ambipolar diffusion, which decouples the neutral and charged particles inside the neutron star (Goldreich \& Reisenegger 1992; Reisenegger 2009), turning the matter from a single, non-barotropic fluid into two weakly interacting fluids. Of these, only the charged fluid will interact with the magnetic field. If it is composed only of protons and electrons, it will be barotropic, thus not stably stratified and much less able (if not completely unable) to sustain hydromagnetic equilibria like those studied here, thus releasing the previously stored energy. If the matter turns into a superfluid/superconducting state, ambipolar diffusion might happen quite quickly (Glampedakis, Andersson \& Lander 2012). The dynamics of this process still remains to be investigated in detail.

Another interesting case is that of Central Compact Objects (CCOs) in supernova remnants. Some of these have been identified as young neutron stars with ``exceptionally weak'' dipole magnetic fields, yet their observed surface temperature anisotropies seem to require the presence of strong magnetic fields hidden in their crusts (Gotthelf, Halpern \& Alford 2013). In particular, one such object, PSR J1852+0040, has a dipole field of a few$\times 10^{10}$\,G. Modeling the surface X-ray emission of this object, Shabaltas \& Lai (2012) concluded that it must possess a toroidal field of strength a few$\times 10^{14}$\,G or larger. This prediction roughly agrees with our limit on the maximal toroidal field strength, which for the surface field in this case indeed gives a few$\times 10^{14}$\,G (equation \ref{max_tor}).

\subsubsection{Comparison with previous numerical simulations}
Our results are in general agreement with the simulations of Braithwaite (2009) for a stably stratified, non-degenerate, polytropic fluid star with $\gamma = 4/3$ (i.e.\ a polytrope of index $n = 3$, which is a reasonable approximation for an upper main-sequence star) and $\Gamma = 5/3$. Braithwaite finds a stability condition of the form $a E_{\rm mag} / U < E_{\rm pol} / E_{\rm mag} \lesssim 0.8$, where $E_{\rm mag} = E_{\rm pol} + E_{\rm tor}$ is the total magnetic energy and $U = E_{\rm grav}/2$ is the thermal energy (by the virial theorem). His simulations yield $a \approx 10$ for main-sequence stars, where $\Gamma/\gamma - 1 \approx 1/4$. Since in realistic stars $E_{\rm mag} / U \ll 1$, this implies that, while the poloidal component cannot be substantially stronger than the toroidal, the toroidal component can be much stronger than the poloidal.

For the lower bound on the poloidal field strength, $E_{\rm pol}$ is small and we have $E_{\rm mag} \approx E_{\rm tor}$. The condition given by equation (\ref{result}) can then be written analogously to Braithwaite as
        \beq
        \frac{E_{\rm pol}}{E_{\rm tor}} \gtrsim 2 a \frac{E_{\rm tor}}{E_{\rm grav}} \ , \mtext{where} a \approx \frac{1.8}{\Gamma/\gamma - 1} \ .
        \label{result2}
        \enq
For main-sequence stars ($\Gamma/\gamma - 1 \approx 1/4$), we obtain $a \approx 7.4$, which compares well with Braithwaite's result of $a \approx 10$. On the other hand, for neutron stars (where, we take $\Gamma/\gamma - 1 \sim 10^{-2}$, which is different than the value $1/400$ used by Braithwaite), we obtain $a \sim 200$.

Notwithstanding the remarkable agreement between the two approaches, they are also notably different. First of all, while our analytic calculations can deal with arbitrary (but small) ratios of magnetic to gravitational energy, Braithwaite is forced to use a specific, and not extremely small value for his simulations ($E_{\rm mag}/U = 1/400$, compared to $\lesssim 10^{-6}$ in real stars). This means that the magnetic force is significantly stronger, and as a result, stable stratification plays a smaller role in stability. This also makes the length scale ratios in the displacement field much less extreme (see equation \ref{frac_delta}), and allows the unstable wavelengths to be resolved, even if only barely. Thus, effectively, the same instabilities should be manifested in both treatments. The scaling of his final result to general values of the ratio $E_{\rm mag}/U$ is then stipulated. Secondly, Braithwaite's grid of values for $E_{\rm pol}/E_{\rm mag}$ has only four values for each configuration, which differ from each other by almost a factor of 2. Therefore, the value of the coefficient $a$ is not much more precise than that. Thirdly, Braithwaite explicitly states that the coefficient $c_m$ (for both $m = 0$ and $m = 1$) is \emph{always} positive for the magnetic fields he considers, and that the instability of the toroidal field results entirely due to the failure of the condition $b_m^2 < 4 a_m c_m$ (equation \ref{conditions}). The first statement seems to expressly contradict our conclusion that for \emph{any} realistic toroidal field, $c_m$ becomes negative at least in some regions, and is the leading source of instability, as pointed out previously by Goossens (1980) and Goossens et al.\ (1981) for various special field configurations. Fourthly, we do not consider the $m = 0$ case in this paper. Tayler (1973) conjectures that the $m = 1$ perturbations ``seem likely to be the worst instabilities in the linear regime'', and Spruit (1999) notes that the $m = 1$ mode ``occurs under the widest range of conditions''. However, as Braithwaite reports, there are instabilities that arise from the $m = 0$ mode as well, and need to be considered. Thus, in our treatment we have only one mode, whereas the numerical simulations in principle have all modes, so the simulations should be more unstable. It is also possible that, with the addition of the poloidal field, the particular displacement field constructed in this paper no longer corresponds to the most unstable mode, which would also imply that we are overestimating the stability. Finally, the hydrostatic background and magnetic field structure are also not identical between the two cases.

\section{Concluding remarks}\label{section_conclusions}
Here, we summarize the main conclusions and discuss further implications of our work.

Stable stratification has an important influence on stellar magnetic equilibria and their stability, by (a) allowing a much larger assortment of possible equilibria, and (b) strongly constraining the displacement fields that might destabilize these equilibria. We can easily construct simple analytic models for axially symmetric magnetic fields compatible with hydromagnetic equilibria in stably stratified stars, with both poloidal and toroidal components of adjustable strengths, as well as the associated pressure, density and gravitational potential perturbations. This makes it possible to directly study their stability. For a weak magnetic field (in the sense that the Alfv\'{e}n frequency is much smaller than the Brunt-V\"{a}is\"{a}l\"{a} buoyancy frequency), the terms in the energy functional involving fluid perturbations due to the magnetic field are small and can be ignored, which makes the algebra much simpler in the cases of poloidal or mixed fields. (For a toroidal field, it does not simplify the algebra much, but it simplifies the physical interpretation.)

There is an important difference between the leading order instabilities of toroidal and poloidal fields. In toroidal fields, instabilities result from the slipping of magnetic loops around the magnetic axis (Tayler 1973). These instabilities are strongly restricted by stable stratification to surfaces of constant radius, but are not completely eliminated. As a result, a relatively weak poloidal field can be sufficient to stabilize the toroidal field (Spruit 1999). On the other hand, in poloidal fields, the instabilities result from the slipping of magnetic loops around the neutral line (Markey \& Tayler 1973; Wright 1973). In this case, stable stratification is of less help in eliminating instabilities, because, while it restricts radial displacements, it does not help with perturbations that are perpendicular to the radial direction, which are just as easily achievable in this case (ignoring curvature effects due to the fact that the neutral line is a circle). Therefore, one might expect that a relatively stronger toroidal field would be needed in order to stabilize a poloidal field. These are consistent with the upper and lower bounds found by Braithwaite (2009).

Previous literature (Tayler 1973; Goossens et al.\ 1981) has given proofs of instability for toroidal fields satisfying special criteria. In particular, it is often repeated that toroidal fields are unstable near the axis (Tayler 1973; Spruit 1999). Here, we prove that in fact all toroidal fields of any realistic structure, in general (barotropic and non-barotropic) fluids, are unstable. This is true even when the toroidal field is contained in a region far away from the axis, as is the case considered in this paper. The instability always happens near the high-latitude limits of the region containing the toroidal field (i.e.\ farthest from the equator and closest to the poles), immaterial of the exact shape of this region. This then allows us to construct a displacement field that should be a reasonable approximation for the most unstable mode, compliant with the constraints of stable stratification.

We find that the toroidal field instability considered in this paper is stabilized by a poloidal field that satisfies equation (\ref{result2}). For main-sequence stars, we find that $a \approx 7.4$, which compares well with the factor of $a \approx 10$ obtained by Braithwaite (2009) through numerical simulations. For neutron stars, we obtain $a \sim 200$. Since observations provide us with an upper limit on the surface poloidal field, this result can then be used to place an upper limit on the internal toroidal field. We find that the energy stored in the toroidal field within the star can be significantly larger than the total energy of the poloidal field, particularly if the latter is weak. Such strong magnetic fields hidden within stars can provide a substantial additional energy budget to power magnetar activity, as well as cause significant stellar distortions with implications for precession and emission of gravitational waves. In particular, implications of the field configuration considered in this paper for gravitational waves are discussed in detail by Mastrano et al.\ (2011).

\section*{Acknowledgements}
This research was supported by FONDECYT Postdoctoral Grant 3085041, FONDECYT Regular Grants 1060644 and 1110213, CONICYT International Collaboration Grant DFG-06, a CONICYT Master's Fellowship, FONDAP Center for Astrophysics (15010003), Basal Center for Astrophysics and Associated Technologies (PFB-06/2007), Proyecto L\'{i}mite VRI-PUC 15/2010, and a Melbourne University International Postgraduate Research Scholarship. We would like to thank Rafael Benguria, Kostas Glampedakis, Maxim Lyutikov, Andrew Melatos, and the anonymous referee for useful discussions and comments.

\appendix
\section{Alternative models for the poloidal field}\label{section_appendix}

        \begin{figure}
        \centerline{\includegraphics[scale=0.46]{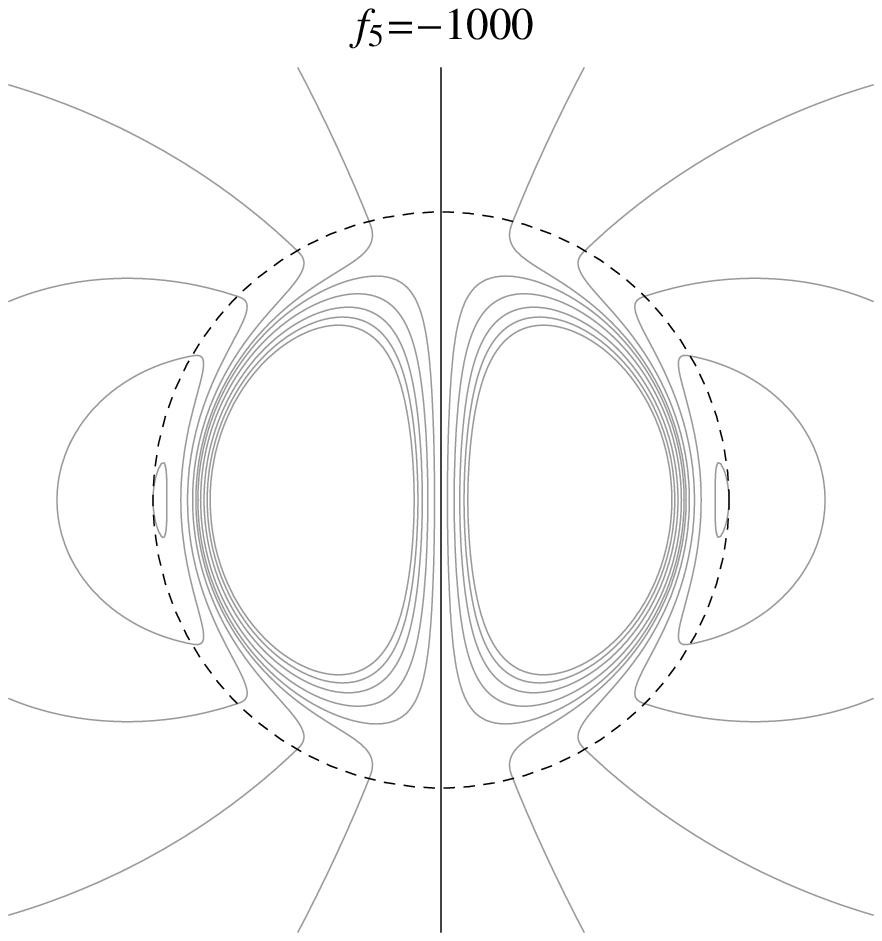} \ \includegraphics[scale=0.46]{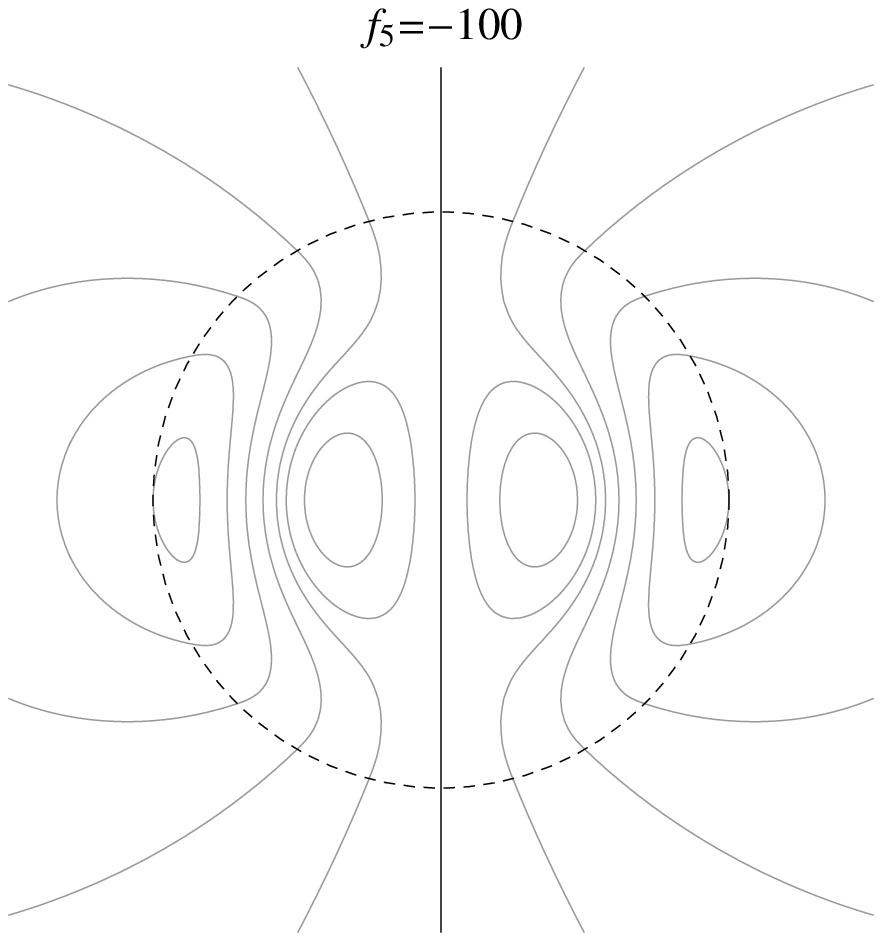} \
        \includegraphics[scale=0.46]{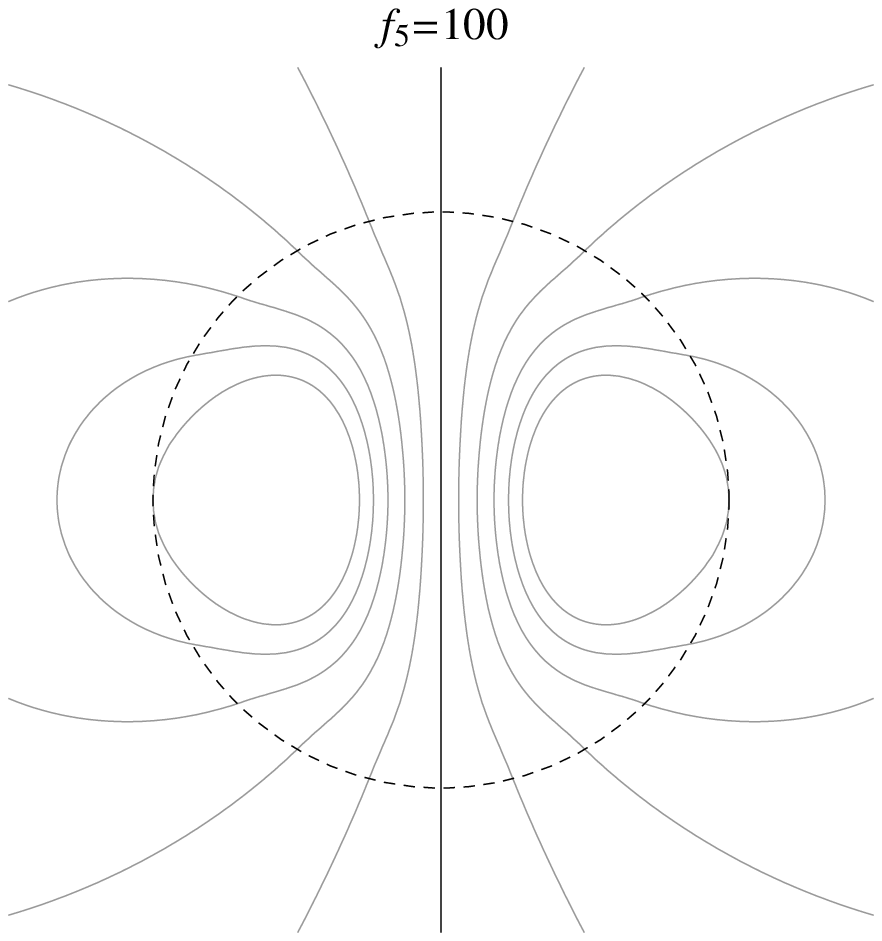} \ \includegraphics[scale=0.46]{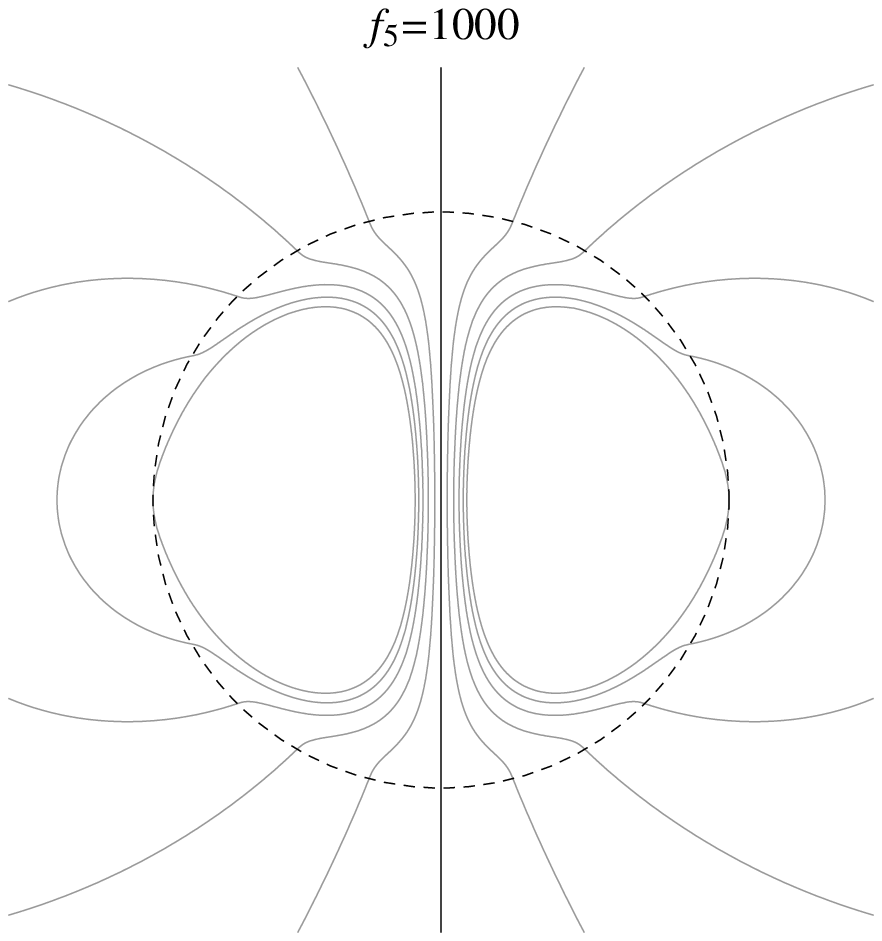}}
        \vspace{0.5cm}
        \centerline{\includegraphics[scale=0.46]{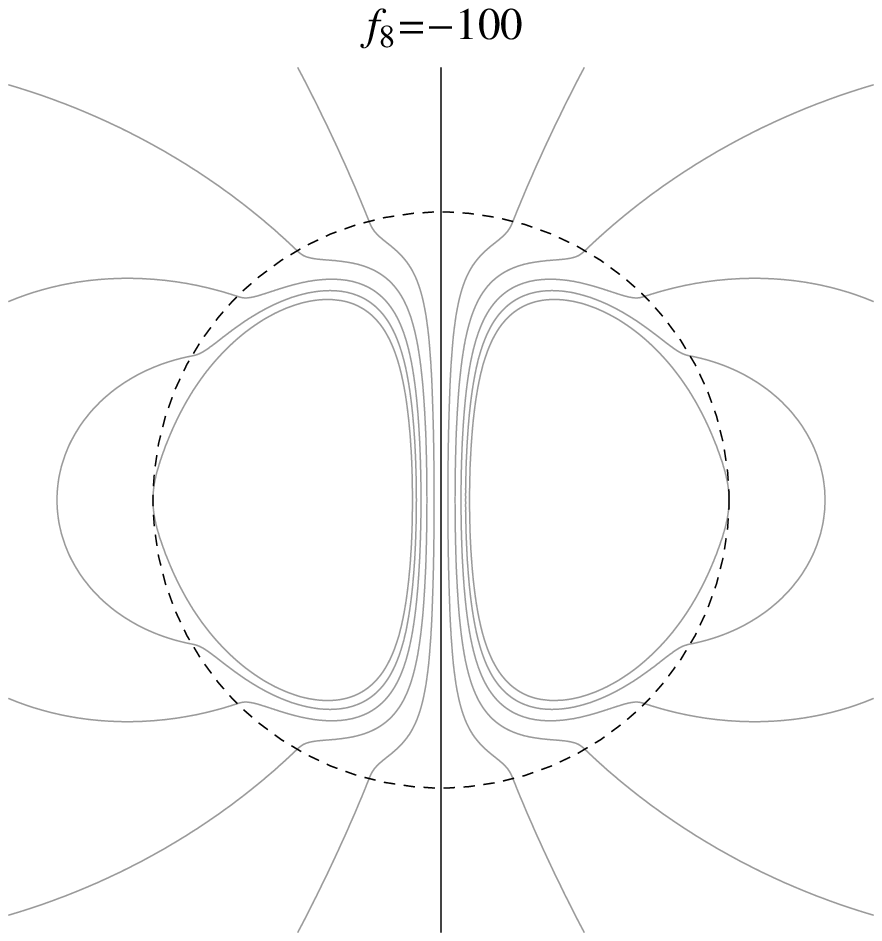} \ \includegraphics[scale=0.46]{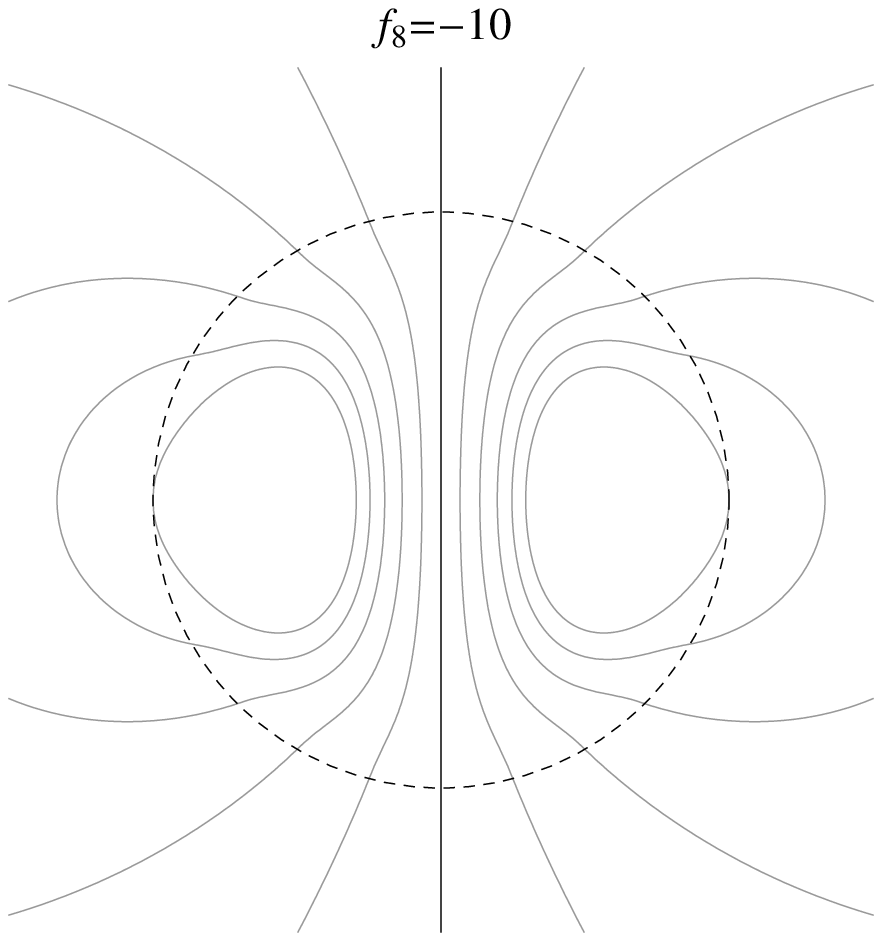} \
        \includegraphics[scale=0.46]{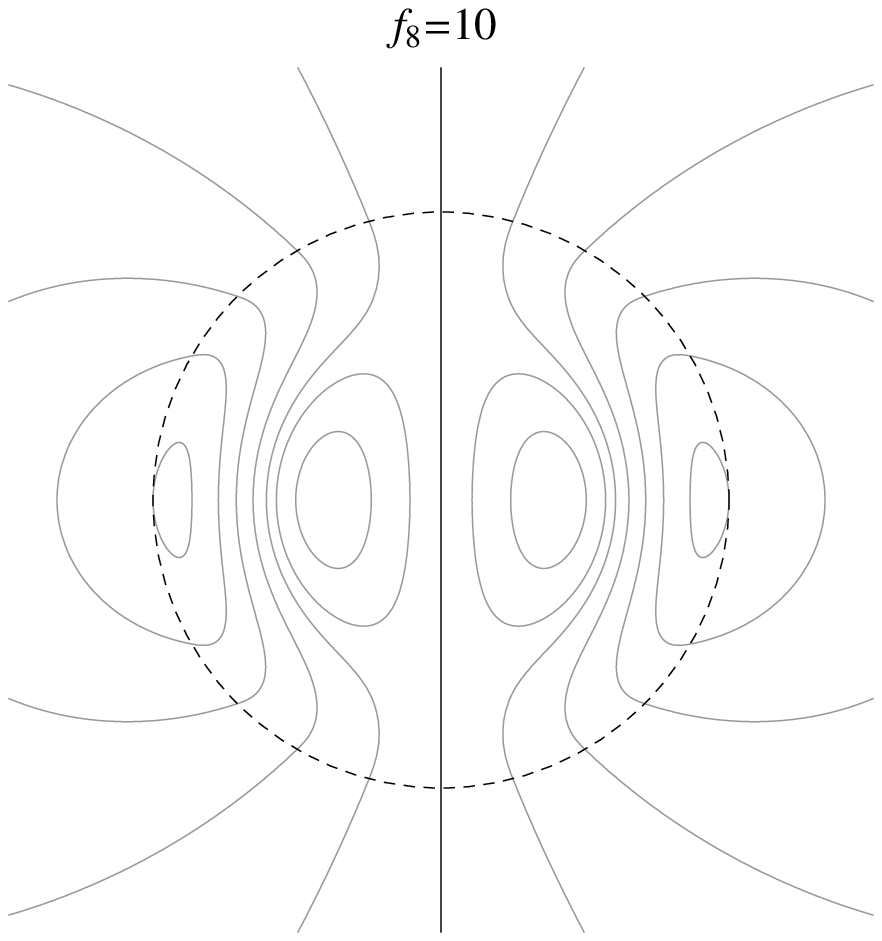} \ \includegraphics[scale=0.46]{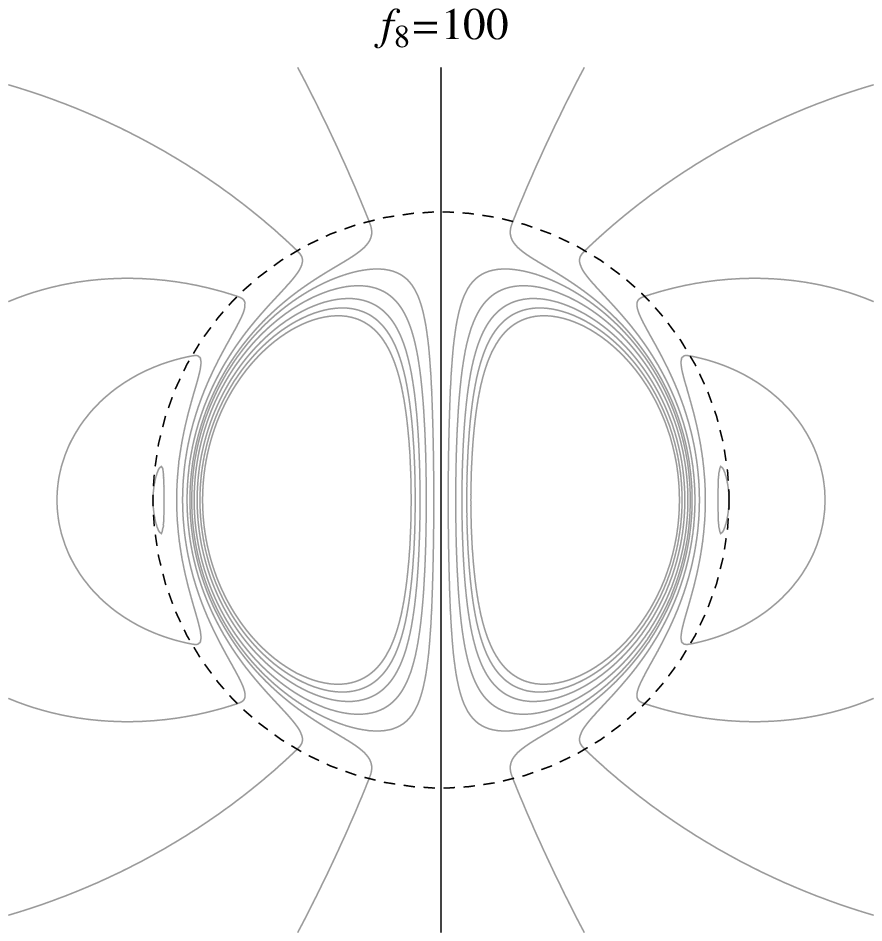}}
        \caption{Poloidal magnetic field lines for various values of the free parameters for the two models given by equations (\ref{model1}) and (\ref{model2}). The stellar surface is shown with a dashed line. The field outside the star is that of a dipole. Four field configurations corresponding to model I are shown in the first line, and four configurations for model II are shown below that. The field configuration of Fig. \ref{fig_field} used throughout the paper is retrieved for $f_5 = 0$ and $f_8 = 0$. The field configuration changes slowly with $f_5$ and $f_8$. For sufficiently positive $f_5$ and for sufficiently negative $f_8$ the region where the toroidal field is present (outlined by the last poloidal field line that closes within the star) grows in size. For sufficiently negative $f_5$ and for sufficiently positive $f_8$, the poloidal field goes through a zero and switches direction somewhere within the star. Thus, a second region of field lines that close within the star is formed, where toroidal fields could also be present.}
        \label{fig_app}
        \end{figure}

Here we discuss alternative models for the poloidal field constructed in \S\ref{section_poloidal}. We still consider cases where the outside field is dipolar, and take the poloidal field function to be of the same form as in equation (\ref{alpha}), i.e.\ $\hat{\alpha}(x,\theta) = f(x)\sin^2\theta$. As before, we are interested in power-law solutions for the radial function of the form $f(x) = \sum f_s x^s$, but we will now somewhat generalize the previous choice of non-zero terms. As discussed in \S\ref{section_poloidal}, regularity conditions imply that we must have either $s = 2$ or $s > 3$. We need at least three terms in order to satisfy the corresponding boundary conditions. In this appendix, we consider the effect of including a fourth term, and construct two model solutions of the form
        \begin{alignat}{2}
        & \mbox{model I:}\hspace{0.6cm} & f(x) & = f_2 x^2 + f_4 x^4 + f_5 x^5 + f_6 x^6 \ ,
        \label{model1} \\
        & \mbox{model II:}\hspace{0.6cm} & f(x) & = f_2 x^2 + f_4 x^4 + f_6 x^6 + f_8 x^8 \ .
        \label{model2}
        \end{alignat}
The first model includes the lowest four allowed terms in the series expansion, and the second model includes the lowest four even terms.

We still need to satisfy the two boundary conditions (equations \ref{boundary1} and \ref{boundary2}), and the normalization condition at the surface,
        \beq
        f'' = \frac{2f}{x^2} \ , \hspace{0.6cm}
        f' = - \frac{f}{x} \ , \mtext{and}
        f = 1 \mtext{at} x = 1 \ .
        \enq
Thus, we have three conditions for a total of four unknowns. Using the three conditions above, we can express three of the coefficients as functions of the remaining free coefficient, which we choose as $f_5$ and $f_8$, respectively. Thus, for the two models, we have
        \begin{alignat}{4}
        & \mbox{model I:}\hspace{0.6cm} & f_2 &= \frac{35}{8} + \frac{f_5}{8} \ , \hspace{0.6cm}
        & f_4 &= - \frac{21}{4} - \frac{3f_5}{4} \ , \hspace{0.6cm} & f_6 &= \frac{15}{8} - \frac{3f_5}{8} \ , \\
        & \mbox{model II:}\hspace{0.6cm} & f_2 &= \frac{35}{8} - f_8 \ , \hspace{0.6cm}
        & f_4 &= - \frac{21}{4} + 3f_8 \ , \hspace{0.6cm} & f_6 &= \frac{15}{8} - 3f_8 \ .
        \end{alignat}
The case considered in \S\ref{section_poloidal} is recovered by setting $f_5 = 0$ and $f_8 = 0$ in the two models, respectively. Sample field configurations for special values of the free coefficients are shown in Fig. \ref{fig_app}. We conclude that: (1) the size of the toroidal region and the strength of the poloidal field within it are monotonically increasing functions of the coefficient $f_5$ and monotonically decreasing functions of the coefficient $f_8$, whichever of the two powers is included. There does not seem to be a limit to this tendency, so for $f_5\to\infty$ or $f_8\to -\infty$ it is likely that $f(x)$ jumps from 0 to 1 right at the origin, so essentially the whole star contains a toroidal field. (2) For a very small toroidal region (very negative $f_5$ or very positive $f_8$) there is a region of negative $f(x)$, i.e.\ oppositely oriented poloidal field lines, around the axis, which could also contain toroidal fields, but may not be realized in actual stars.


\begin{thebibliography}{}
\bibitem[Akg\"{u}n \& Wasserman 2008]{akgunref}
        Akg\"{u}n T., Wasserman I., 2008, MNRAS, 383, 1551
\bibitem[Bernstein et al. 1958]{bernsteinref}
        Bernstein I. B., Frieman E. A., Kruskal M. D., Kulsrud R. M., 1958, Proc. R. Soc. A, 244, 17
\bibitem[Braithwaite 2009]{braithwaiteref1}
        Braithwaite J., 2009, MNRAS, 397, 763
\bibitem[Braithwaite \& Spruit 2004]{braithwaiteref2}
        Braithwaite J., Spruit H., 2004, Nature, 431, 819
\bibitem[Braithwaite \& Nordlund 2006]{braithwaiteref3}
        Braithwaite J., Nordlund {\AA}., 2006, A\&A, 450, 1077
\bibitem[Chandrasekhar 1981]{chandraref}
        Chandrasekhar S., 1981, \emph{``Hydrodynamic and Hydromagnetic Stability''}, Dover, New York
\bibitem[Chandrasekhar \& Fermi 1953]{chandraref2}
        Chandrasekhar S., Fermi E., 1953, ApJ, 118, 116
\bibitem[Chandrasekhar \& Prendergast 1956]{chandraref3}
        Chandrasekhar S., Prendergast K. H., 1956, Proc. Nat. Acad. Sci., 42, 5
\bibitem[Dicke 1979]{dickeref}
        Dicke R. H., 1979, ApJ, 228, 898
\bibitem[Ferraro 1954]{ferraroref}
        Ferraro V. C. A., 1954, ApJ, 119, 407
\bibitem[Flowers \& Ruderman 1977]{flowersref}
        Flowers E., Ruderman M. A., 1977, ApJ, 215, 302
\bibitem[Friedman \& Schutz 1978]{friedmanref}
        Friedman J. L., Schutz B. F., 1978, ApJ, 221, 937
\bibitem[Glampedakis, Andersson \& Lander 2012]{glampedakisref}
        Glampedakis K., Andersson N., Lander S. K., 2012, MNRAS, 420, 1263
\bibitem[Goldreich \& Reisenegger 1992]{goldreichref}
        Goldreich P., Reisenegger A., 1992, ApJ, 395, 250
\bibitem[Goossens 1980]{goossensref1}
        Goossens M., 1980, Geophys. Astrophys. Fluid Dynamics, 15, 123
\bibitem[Goossens \& Veugelen 1978]{goossensref2}
        Goossens M., Veugelen P., 1978, A\&A, 70, 277
\bibitem[Goossens \& Biront 1980]{goossensref3}
        Goossens M., Biront D., 1980, SSRv, 27, 667
\bibitem[Goossens \& Tayler 1980]{goossensref4}
        Goossens M., Tayler R. J., 1980, MNRAS, 193, 833
\bibitem[Goossens et al. 1981]{goossensref5}
        Goossens M., Biront D., Tayler R. J., 1981, Ap\&SS, 75, 521
\bibitem[Gotthelf et al. 2013]{gotthelfref}
        Gotthelf E. V., Halpern J. P., Alford J., 2013, ApJ, 765, 58
\bibitem[Haskell et al. 2008]{haskellref}
        Haskell B., Samuelsson S., Glampedakis K., Andersson N., 2008, MNRAS, 385, 531
\bibitem[L\"{u}st \& Schl\"{u}ter 1954]{lustref}
        L\"{u}st R., Schl\"{u}ter A., 1954, Zs. f. Ap., 34, 263
\bibitem[Marchant et al. 2011]{marchantref}
        Marchant P., Reisenegger A., Akg\"{u}n T., 2011, MNRAS, 415, 2426
\bibitem[Markey \& Tayler 1973]{markeyref}
        Markey P., Tayler R. J., 1973, MNRAS, 163, 77
\bibitem[Mastrano et al. 2011]{mastranoref}
        Mastrano A., Melatos A., Reisenegger A., Akg\"{u}n T., 2011, MNRAS, 417, 2288
\bibitem[Mereghetti 2008]{mereghettiref}
        Mereghetti S., 2008, A\&AR, 15, 225
\bibitem[Mestel 1956]{mestelref}
        Mestel L., 1956, MNRAS, 116, 324
\bibitem[Prendergast 1956]{prendergastref}
        Prendergast K. H., 1956, ApJ, 123, 498
\bibitem[Rea et al. 2010]{rearef}
        Rea N. et al., 2010, Sci, 330, 944
\bibitem[Reisenegger 2009]{reiseneggerref}
        Reisenegger A., 2009, A\&A, 499, 557
\bibitem[Shabaltas \& Lai 2012]{shabaltasref}
        Shabaltas N., Lai D., 2012, ApJ, 748, 148
\bibitem[Spruit 1999]{spruitref}
        Spruit H. C., 1999, A\&A, 349, 189
\bibitem[Tayler 1973]{taylerref}
        Tayler R. J., 1973, MNRAS, 161, 365
\bibitem[Thompson \& Duncan 2001]{thompsonref}
        Thompson C., Duncan R. C., 2001, ApJ, 561, 980
\bibitem[Wright 1973]{wrightref}
        Wright G. A. E. 1973, MNRAS, 162, 339
\end{thebibliography}
\end{document}